\documentclass[aps,prc,
showpacs,twocolumn,superscriptaddress,nofootinbib]{revtex4-1}

\usepackage{graphicx}
\usepackage{colordvi}
\usepackage{color}
\usepackage{bm}
\usepackage{amsmath}
\usepackage{amssymb}
\usepackage{mathtools}
\usepackage{verbatim}
\usepackage{multirow,array}
\usepackage{bigstrut}
\usepackage{array}
\usepackage{fix-cm}
\usepackage{bbold}
\usepackage{booktabs}
\usepackage{subfigure}

\newcommand{\mU}{\mathcal{U}}
\newcommand{\mV}{\mathcal{V}}

\newcommand{\ii}{\textrm{i}}

\def\a{\alpha}
\def\b{\beta}
\def\e{\varepsilon}

\def\d{\delta}
\def\g{\gamma}
\def\m{\mu}

\def\om{\omega}

\def\ran{\rangle}

\def\lan{\langle}

\def\ii{\textrm{i}}

\newcommand{\ad}[1]{a_{#1}}
\newcommand{\ac}[1]{a^{\dagger}_{#1}}
\newcommand{\bra}[1]{\langle #1 \vert}
\newcommand{\ket}[1]{\vert #1 \rangle}

\newcommand{\be}{\begin{equation}}
\newcommand{\ee}{\end{equation}}
\newcommand{\bea}{\begin{eqnarray}}
\newcommand{\eea}{\end{eqnarray}}

\def\ran{\rangle}
\def\lan{\langle}
\def\a{\alpha}
\def\b{\beta}

\begin{document}

\title{Chiral three-nucleon forces and the evolution of correlations \\ along the oxygen isotopic chain}

\author{A. Cipollone}
\affiliation{Department of Physics, University of Surrey, Guildford GU2 7XH, UK}
\affiliation{Numerical Applications and Scenarios Division, CMCC, Bologna 40127, Italy}

\author{C. Barbieri}
\email{c.barbieri@surrey.ac.uk}
\affiliation{Department of Physics, University of Surrey, Guildford GU2 7XH, UK}

\author{P. Navr\'atil}
\affiliation{TRIUMF, 4004 Westbrook Mall, Vancouver, BC, V6T 2A3, Canada}

\date{\today}

\begin{abstract}

\begin{description}

\item[Background]
 Three-nucleon forces (3NFs) have non trivial implications on the evolution of correlations at extreme proton-neutron asymmetries. Recent {\em ab initio} calculations show that leading order chiral interactions are crucial to obtain the correct binding energies and neutron driplines along the O, N, and F chains, Ref.~[A. Cipollone, C. Barbieri, P. Navr\'atil, Phys.\ Rev.\ Lett \textbf{111}, 062501 (2013)].
\item[Purpose]
 Here, we discuss the impact of 3NFs along the oxygen chain for other quantities of interest, such has the spectral distribution for attachment and removal of a nucleon, spectroscopic factors and radii. The objective is to better delineate the general effects of 3NFs on nuclear correlations.
\item[Methods]
We employ self-consistent Green's function (SCGF) theory which allows a comprehensive calculation of the single particle spectral function. For the closed subshell isotopes, $^{14}$O, $^{16}$O, $^{22}$O, $^{24}$O and $^{28}$O, we perform calculations with the Dyson-ADC(3) method which is fully non-perturbative and is the state of the art for both nuclear physic and quantum chemistry applications. The remaining open shell isotopes are studied using the newly developed Gorkov-SCGF formalism up to second order.
\item[Results]
We produce complete plots for the spectral distributions. The spectroscopic factors for the dominant quasiparticle peaks are found to depend very little on the leading order (NNLO) chiral 3NFs. The latters have small impact on the calculated matter radii, which, however are  consistently obtained smaller than experiment. Similarly, single particle spectra tend to be too spread with respect to the experiment. This effect might hinder, to some extent, the onset of correlations and screen the quenching of calculated spectroscopic factors. The most important effect of 3NFs is thus the fine tuning of the energies for the dominant quasiparticle states, which governs the shell evolution and the position of driplines.
\item[Conclusions]
Although present chiral NNLO 3NFs interactions do reproduce the binding energies correctly in this mass region, the details of the nuclear spectral function remain at odd with the experiment showing too small radii and a too spread single particle spectrum, similar to what already pointed out for larger masses.  This suggests a lack of repulsion in the present model of NN+3N interactions which is mildly apparent already for masses in the A=14--28 range.
\end{description}

\end{abstract}

\maketitle


\section{Introduction}

 The concept of correlations is fundamental to a deep understanding of nuclear phenomena~\cite{Dickhoff2004ppnp}. These are generally defined as characteristics of the nucleus that cannot be explained in terms of a simple mean field picture (i.e., a wave function of Slater determinant type). These effects are often quantified in terms of the fragmentation of the single particle strength observed when  adding or removing a nucleon.  An intriguing feature is the persistence of dominant quasiparticle peaks near the Fermi surface while broader resonances are found at higher excitations. This is at the origin of the duality between the liquid drop and the shell model behaviour of atomic nuclei.

Historically, several electron scattering studies have provided a wealth of information on nuclear spectral functions (see \cite{Mougey1976eep,denHerder1988ZrV,Lapikas1993eep,Onderwater1997prl,Rohe2004prl,Barbieri2004rescatt,Subedi2008sci,Middleton:2006O16pn} and references therein). This has allowed a rather complete characterisation of correlations for stable nuclei~\cite{Dickhoff2004ppnp}.
However, a similar full characterization for exotic isotopes is still lacking.
Recent data from radioactive beam facilities have put in evidence new phenomena such as shell evolution with changing proton-neutron asymmetry~\cite{Schiffer2004prl} and the insurgence of new magic numbers~\cite{Janssens2009natureO24,Wienholtz:2013nya,Steppenbeck2013nature}. 
From the theoretical point of view, some of these effects have been explained in terms of properties of the tensor interaction~\cite{Otsuka2005prl}, and the need for contributions from three-nucleon forces (3NFs) has also been pointed out~\cite{Zuker2003prl}. 
More recently, it has been shown that 3NFs are crucial for understanding the neutron rich side of the nuclear chart. In particular, to explain the oxygen dripline at $^{24}$O~\cite{Otsuka2010prl,Hagen2012prlOx,Hergert2013prl,Cipollone2013prl} and neutron rich Ca isotopes~\cite{Holt2011jpg,Hagen2012prlCa,Soma2014s2n,Hergert2014Ni}.
Ref.~\cite{Cipollone2013prl} found that the same mechanism responsible for the anomalous oxygen dripline also affects N and F isotopes up to at least $^{29}$F, which is strongly neutron rich but still not at the dripline.

{\em Ab initio} calculations of atomic nuclei have advanced dramatically in the medium mass region. Several approaches such as coupled cluster~\cite{Hagen2014CCMrev, Binder2014ccSn132}, in-medium similarity renormalisation group (IM-SRG)~\cite{Tsukiyama2011prl} and self-consistent Green's function (SCGF)~\cite{Barbieri2009ni,Cipollone2013prl} theories are now capable of approaching masses up to A$\approx$100 or more.
These allow to fully exploit modern chiral interactions with two-nucleon (NN) and 3NFs evolved through SRG techniques~\cite{Jurgenson2009prl}. 
Moreover, open shell nuclei have become accessible through a Gorkov extension of the SCGF formalism~\cite{Soma:2011GkvI,Soma:2013rc,Soma2014s2n}, multi-reference IM-SRG~\cite{Hergert2013prl,Hergert2014Ni} and Bogoliubov coupled cluster (BCC)~\cite{SignoracciBCC}.
These \hbox{\em ab initio} studies have mainly focussed on ground state properties, such as total binding energies and two-nucleon separation energies.
More recent works have addressed the construction of effective shell model interactions directly from full NN plus 3NF Hamiltonians~\cite{Bogner2014prlOxSM,Jansen2014prlOxSM}. This allows to successfully address low-energy excitations directly from first principles. 

The SCGF method has the added advantage to provide consistent optical potentials and spectral functions over the whole energy spectrum (i.e., both close and far form the Fermi surface). This gives comprehensive insights into the many-body dynamics and allows to address other quantities such as giant resonances or the qualitative features of single particle distribution~\cite{Mougey1976eep,Mougey1980eep}, which can require
considering several major shells~\cite{Barbieri2009prl,Barbieri2009ni}.

In this paper, we consider the Green's functions of the oxygen isotopes already obtained in Ref.~\cite{Cipollone2013prl} and extend these calculations to the remaining even mass---and open shell---isotopes, using the Gorkov-SCGF approach.
We then present first fully microscopic calculations of the evolution of the single particle spectral functions along a full isotopic chain. This gives an overall description of the evolution of nuclear correlations between two extremes of the nuclear chart. At the same time, it allows to perform a more thorough test of modern chiral interactions and, in particular, we investigate the effects of initial 3NFs at NNLO.

Section~\ref{formalism} discusses the relevant features of the SCGF formalism, for completeness. It reviews the links of propagators with the spectral function and other quantities of experimental interest. Calculations are done in an {\em ab initio} fashion and  we discuss in some details the choice of the Hamiltonian, the approximations taken and the expected uncertainties, when these can be estimated. This is done in Sec.~\ref{calcdtls}.
Sec.~\ref{results} discusses our results for single particle spectra, spectroscopic factors and binding energies. Full three-dimensional plots of spectral functions are discussed in the Appendix for completeness and conclusions are drawn in Sec.~\ref{conlusions}.

\section{SCGF formalism}
\label{formalism}

Information about the single particle dynamics is fully contained in the one-body Green's function, or propagator, whose Lehmann representation reads: 
\begin{align}
 g_{\alpha \beta}(\omega) ~=~& 
 \sum_n  \frac{ 
          \bra{\Psi^A_0}  	\ad{\a}   \ket{\Psi^{A+1}_n}
          \bra{\Psi^{A+1}_n}  \ac{\b}  \ket{\Psi^A_0}
              }{\omega - \e_n^{+} + \ii \eta }  ~+~\nonumber\\
 ~+~ &\sum_k \frac{
          \bra{\Psi^A_0}      \ac{\b}    \ket{\Psi^{A-1}_k}
          \bra{\Psi^{A-1}_k}  a_{\a}	    \ket{\Psi^A_0}
             }{\omega - \e_k^{-} - \ii \eta } \; .
\label{eq:g1}
\end{align}
In Eq.~\eqref{eq:g1}, $\ket{\Psi^{A}_0}$ represents the ground state of A nucleons and $\ket{\Psi^{A+1}_n}$, $\ket{\Psi^{A-1}_k}$ are the eigenstates  of the ($A\pm1$)-nucleon system. The greek indices $\alpha$,$\beta$,..., label a complete orthonormal single particle basis, while
\hbox{$\varepsilon_n^{+}\equiv(E^{A+1}_n - E^A_0)$} and 
\hbox{$\varepsilon_k^{-}\equiv(E^A_0 - E^{A-1}_k)$}
are one-nucleon addition and removal energies, respectively. 
 Note that these are generically referred to in the literature as ``separation'' or ``quasiparticle'' energies although the first naming normally refers to transitions involving only ($A\pm1$)-nucleon ground states.  We will use the second convention in the following, unless the two naming are stricktly equivalent.
The transition amplitudes
\hbox{${\cal X}^n_{\a}\equiv\lan\Psi_n^{A+1}|a_{\a}^{\dag}|\Psi_0^A\ran$} and
\hbox{${\cal Y}^k_{\a}\equiv\lan\Psi_k^{A-1}|a_{\a}|\Psi_0^A\ran$} give information about the strength of the corresponding particle addition and removal processes.

The  one-body Green's function~\eqref{eq:g1} is completely determined by solving the Dyson equation,
\begin{equation}
  \label{eq:Dy}
  g_{\a\b}(\om)=g^{0}_{\a\b}(\om)+ \sum_{\g\d}g^{0}_{\a\g}(\om)\Sigma_{\g\d}^{\star}(\om)g_{\d\b}(\om)  \; ,
\end{equation}
 where the unperturbed propagator $g^{0}_{\a\b}(\om)$ is the initial reference state (usually a mean field or Hartree-Fock state) while $g_{\a\b}(\om)$ is the correlated propagator. 
A full knowledge of the self-energy  $\Sigma_{\a\b}^{\star}(\om)$ yields the exact solution for $g_{\a\b}(\om)$. However, in practical calculations this has to be approximated and it is expanded in terms of the propagator itself (that is, $\Sigma^{\star}=\Sigma^{\star}[g(\om)]$). Thus, an iterative procedure is required to solve for $\Sigma^{\star}(\om)$ and Eq.~\eqref{eq:Dy} self consistently. The approximation schemes we employ to calculate the self-energy are outlined in the next subsection. 

The attractive feature of the SCGF approach is that $g_{\a\b}(\om)$ describes the one-body dynamics completely. The particle and hole spectral functions are extracted directly from Eq.~\eqref{eq:g1}, respectively:
 \begin{eqnarray}
\nonumber 
S^{p}_{\a \b}(\omega)&=&\sum_{n} \left({\cal X}^n_\a \right)^*  {\cal X}^n_\b ~ \delta\Big(\omega-(E^{A+1}_n - E^A_0)\Big) \; ,
\\
 \label{eq:SpSh}
S^{h}_{\a \b}(\omega)&=&\sum_{k}  {\cal Y}^k_\a  \left({\cal Y}^k_\b \right)^* ~\delta\Big(\omega-(E^A_0 - E^{A-1}_k)\Big) \; .
\end{eqnarray} 
Any one-body observable can be calculated via the one-body density matrix $\rho_{\a\b}$, which is obtained from $g_{\a\b}$ as follows:
\begin{align}
  \label{eq:rho}
  \rho_{\alpha \beta}&\equiv\bra{\Psi_0^A}a^{\dag}_{\b}a_{\alpha} \ket{\Psi_0^A}\nonumber\\
  &=\int_{-\infty}^{\e^-_0}S^h_{\a\b}(\om)~d\om=\sum_k ({\cal Y}^k_{\b})^*{\cal Y}^k_{\a} .
\end{align}
The expectation value of a one-body operator, ${\hat O}^{1B}$, can then be written in terms of the ${\cal Y}$ amplitudes as:
\begin{equation}
\label{eq:den_one}
\langle {\hat O}^{1B}\rangle  =\sum_{\a\b}  O^{1B}_{\a \b}\,\,\rho_{\b\a}=\sum_k \sum_{\a\b}~ ({\cal Y}^k_{\a})^*~ O^{1B}_{\alpha,\b} ~  {\cal Y}^k_{\b} \; .
\end{equation}
Evaluating two- and many-nucleon observables requires the knowledge of many-body propagators. In the following, we do this by approximating the corresponding A-body density matrices with A correlated but non-interacting propagators, Eq.~\eqref{eq:rho}.
Specifically, we use this to account for the centre of mass (COM) correction when calculating root mean square (rms) radii:
\begin{align}\label{eq:r_rms}
\lan r^2\ran&\simeq\frac{1}{A}\sum_{\a\b} u_\alpha \, \lan\a| \vec{\bf r}\,^2|\b\ran~\rho_{\b\a}~\nonumber\\
&~-\frac{1}{A^2}\sum_{\a\b\g\d} u_{\alpha\gamma} \, \lan\a\g|\vec{\bf r}_1\cdot\vec{\bf r}_2|\b\d\ran ~\rho_{\b\a}\rho_{\d\g} \; ,
\end{align}
where $\vec{\bf r}_i$ represents the position of particle $i$. 
The factors $u_\alpha$ and $u_{\alpha \beta}$ in Eq.~\eqref{eq:r_rms} and the two-body correction term arise because the intrinsic radius is calculated with respect to the COM of the system~\cite{Hergert:2009na}. Point-matter radii are calculated by taking $u^{pt-m}_\alpha=(A-1)/A$ and $u^{pt-m}_{\alpha \gamma}=1$, while point-proton radii are found using
\begin{equation}
  \label{eq:ua_chrg}
  u^{pt-p}_\alpha = \left\{
  \begin{array}{cll}
     \frac{A(A-2)+Z}{ZA} & & \textrm{if $\alpha$ labels a proton state} ,\\
     ~\\
     \frac{1}{A} & & \textrm{if $\alpha$ is a neutron}  
  \end{array}  \right.
\end{equation}
and
\begin{equation}
  \label{eq:uab_chrg}
  u^{pt-p}_{\alpha\beta} = \left\{
  \begin{array}{cll}
     \frac{2A-Z}{Z} & & \textrm{if $\alpha$,$\gamma$ label  two proton states}  , \\
     ~\\
     \frac{A-Z}{Z} & & \textrm{if $\alpha$,$\gamma$ are a proton and a neutron}  , \\
     ~\\
     -1 & & \textrm{if $\alpha$,$\gamma$ are two neutrons}   .
  \end{array}  \right.
\end{equation}
To obtain charge radii, we first calculate the point-proton ones and then account for 
the rms charge radii of the nucleons and for the Darvin-Foldy 
relativistic correction~\cite{Friar1997DwFld}:
\begin{equation}
\lan r^2_{ch}\ran = \lan r^2_{pt-p}\ran + \lan R_p^2 \ran + \frac{N}{Z}\lan R_n^2 \ran + \frac{3\hbar^2}{4m_p^2c^2} \; ,
\end{equation}
with  $\lan R_p^2 \ran$=0.8775(51)~fm${}^2$~\cite{Mohr2012rmpCODATA} and $\lan R_n^2 \ran$=$-$0.1149(27) fm${}^2$~\cite{Angeli2013ADNDT}.
 In the present calculations, the contribution of second term of Eq.~\eqref{eq:r_rms} to the rms radii are 
$\leq$~0.03~fm and decrease with the mass number. Refs.~\cite{Cipollone2013prl,Barbieri2014qmbt17} have considered
first order corrections to the approximation
of  A non interacting propagators---used to calculate this term---and found that it is negligible in most cases as long as fully correlated densities
are used. Therefore, we conclude that Eq.~\eqref{eq:r_rms} does not introduce sizable errors.

The exact one-body propagator, $g_{\a\b}(\om)$, also allows calculating the total energy by means of the extended Koltun sum-rule \cite{Carbone2013tnf}: 
\begin{equation}
  \label{eq:Koltun_hW}
  E^A_0 = \sum_{\alpha\beta} \frac{1}{2} 
            \int_{-\infty}^{\e^-_0}  [\,T_{\alpha\beta}+\omega\,\delta_{\alpha\beta}\, ]
            S^h_{\b\a}(\omega) d\om
            -  \frac{1}{2} \langle W\rangle \, .
\end{equation}
This requires only the additional evaluation of the expectation value of the three-nucleon interaction,~$\lan W \ran$. Again, we approximate this in terms of non-interacting three-body density matrices:
\begin{equation}
   \label{eq:Wddd}
    \lan W\ran\simeq\frac{1}{6} \, \sum_{\a\b\mu\g\d\nu} W_{\a\b\mu,\g\d\nu}~\rho_{\g\a}~\rho_{\d\b}~\rho_{\nu\mu} \; .
\end{equation}
The errors in this approximation have been estimated in Ref.~\cite{Cipollone2013prl} and were found to not exceed the 250~keV on the total binding energy for $^{16}$O and $^{24}$O. 

In all simulations below we subtract the spurious contribution of the  kinetic energy of the COM and work with the intrinsic Hamiltonian \hbox{$H[A]=H-T_{COM}(A)=U(A)+V(A)+W$}, which acquires a dependence on total  number of nucleons. The $U$, $V$ and $W$ label  one-, two- and three-body interactions. This implies that the particle and hole spectra of the even-odd isotopes  are recalculated separately from ${H}[A +1]$ and ${H}[A - 1]$. They are then corrected for the COM motion as follows:
\begin{equation}
\begin{array}{l}
\e^{+}_{n,\textrm{COM}}= \e^{+}_n[A+1]+E_0^A[A+1]-E_0^A[A]\vspace{0.3cm}  \; , \\
\e^{-}_{k,\textrm{COM}}= \e^{-}_k[A-1]-E_0^A[A-1]+E_0^A[A]  \; ,
\end{array}
\label{eq:com}
\end{equation}
where  $\e^{\pm}_n[A\pm 1]$ and $E^{A}_k[A\pm 1]$ label the poles of  $g_{\a\b}(\om)$ and the total energies, Eq.~\eqref{eq:Koltun_hW}, calculated from the \hbox{${H}[A\pm 1]$} Hamiltonian. The overall COM corrections become progressively smaller as $A$ increases.

\subsection{Dyson-ADC(3) and second-order Gorkov equations}

Calculations with 3N interactions follows the procedure extensively discussed in Ref.~\cite{Carbone2013tnf}, which involves defining the following medium dependent one- and two-body interactions:
\begin{align}
  \label{eq:UV_eff}
  &\widetilde{U}_{\a\b}=U_{\a\b} + \sum_{\d\g} V_{\a\g,\b\d}~\rho_{\d\g}+ \frac{1}{4} \sum_{\mu \nu \g \d}W_{\a\mu\nu,\b\g\d}~\rho_{\g\mu}~\rho_{\d\nu}  \; ,
  \nonumber\\
  &\widetilde{V}_{\a\b,\g\d}=V_{\a\b,\g\d}+\sum_{\mu\nu}W_{\a\b \mu ,\g \d \nu}~\rho_{\nu\mu}  \; .
\end{align}
 This allows neglecting residual contributions in $W$ that have been found to be negligible  for oxygen isotopes~\cite{Hagen2007cc3nf,Roth2012prl}. 
 Hence, we retain only interaction-irreducible diagrams in $\widetilde{U}$ and $\widetilde{V}$ to the self-energy.

To solve Eq.~\eqref{eq:Dy}, we express the self-energy as,
\begin{align}\label{eq:ADCn}
\Sigma_{\a\b}^{\star}(\omega)=&\Sigma_{\a\b}(\infty)+\sum_{i \, j}{\bm D}^{\dag}_{\a i}\left[\frac{1}{\omega-({\bm K}+{\bm C})}\right]_{i \, j}{\bm D}_{j \b} ~.
\end{align}
where $\Sigma_{\a\b}(\infty)$ is the correlated and energy-independent mean field. 
The whole $\Sigma^*_{\a\b}(\om)$ is an  optical potential for elastic scattering of a nucleon off the $|\Psi^A_0\ran$ ground state, which also describes the fragmentation of the particle and hole spectra~\cite{Capuzzi96,Escher2002opt}.

In Eq.~\eqref{eq:ADCn}, the matrix {\bf D} couples single particle states to more complex intermediate configurations, while {\bf K} and {\bf C} are their unperturbed energies and interaction matrices. For the closed subshell isotopes we exploit the third order algebraic diagrammatic construction  [ADC(3)] scheme, which is the best compromise between computational efforts and accuracy. This consists in the minimal choice of these matrices that retains all self-energy diagrams up to third order. Although ADC(3) is constrained at third order, it contains infinite order summations of diagrams that include particle-particle and hole-hole ladders as well as particle-hole rings. It is therefore a fully non-perturbative approach.
Generally speaking, ADC($n$) defines a hierarchy of truncation schemes of Eq.~\eqref{eq:ADCn} for increasing order $n$ that allows systematic improvements of the method~\cite{Schirmer1983ADCn}.

Recently, SCGF theory has been extended to a Nambu-Gorkov formulation that allows  addressing truly open shell nuclei~\cite{Soma:2011GkvI}. This has opened the possibility to calculate ground state properties and the one-nucleon addition/removal spectra of mid-mass open shell nuclei, in a fully {\em ab initio} fashion.
As in BCS theory, one allows for an explicit breaking of particle-number conservation that is necessary for a proper description of \emph{pairing} correlations~\cite{Soma:2011GkvI,Idini2012}.
This implies introducing a grand canonical Hamiltonian $\Omega=H-\m_n \hat{N}-\m_p \hat{Z}$ and constraining the proton (neutron) chemical potentials $\m_p(\m_n)$ to recover the correct particle number on average: $A=\bra{\Psi_0}\hat{A}\ket{\Psi_0}$, where $\ket{\Psi_0}$ is the symmetry-broken ground state. A detailed description of the theory can be found in Refs.~\cite{Soma:2011GkvI,Soma2014GkvII,Soma:2013rc}.

In Gorkov theory one is left with a set of {\em normal} and {\em anomalous} propagators and self-energies with similar Lehman representations to Eqs.~\eqref{eq:g1} and~\eqref{eq:ADCn}. In particular, the normal propagator is
\begin{eqnarray}
\label{eq:Gab11}
G^{11}_{\a \b} (\om) &=&  \sum_{k} \left\{
\frac{\mU_{\a}^{k} \,\mU_{\b}^{k*}}
{\om-\om_{k} + \ii \eta}
+ \frac{\bar{\mV}_{\a}^{k*} \, {\bar{\mV}_{\b}^{k}}}{\om+\om_{k} - \ii \eta} \right\} \: 
\end{eqnarray}
where $\mU,\mV$ are the transition amplitudes for reaching the states $| \Psi_{k} \rangle$  by adding (removing) a nucleon to (from) $| \Psi_{0} \rangle$, and $\omega_{k}$ are the corresponding quasiparticle energies~\cite{Soma:2011GkvI}. 

 The Gorkov version of the SCGF approach allows to calculate spectral functions for open shell semi-magic systems.
Its present formulation follows the ADC($n$) truncation scheme discussed above but has been implemented only up
to the second order.  This has allowed successful predictions of trends in binding
energies~\cite{Soma2014s2n}. However, the ADC(2) is known to slightly underestimate binding energies and it is not guaranteed
to provide accurate predictions for one-nucleon removal and addition energies~\cite{vonNiessen1984CPR}, which are instead possible
with a Dyson-ADC(3) calculations. The full extension to Gorkov-ADC(3) formalism is currently
underway~\cite{GADC3}.

\section{Calculations}
\label{calcdtls}

Calculations have been performed using NN and 3N chiral interactions evolved to a low-momentum scale $\lambda_{SRG}$ through free space similarity renormalization group~(SRG) techniques~\cite{Jurgenson2009prl}. The original NN interaction is the next-to-next-to-next-to-leading order (N3LO) with a cutoff $\Lambda_{\textrm{NN}}=500$ MeV/c, from Refs.~\cite{Entem2003N3LO,Machleidt2011pr}. For the 3N interactions we used the NNLO with a reduced local cutoff of $\Lambda_{\textrm{3N}}=400$ MeV/c~\cite{Navratil2007tnf,Roth2012prl}. This includes the two-pion exchange contribution that was originally proposed by Fujita and Miyazawa~\cite{FujitaMiyazawa19573nf}.  Low-energy constants were set at $c_D=-0.2$, $c_E=0.098$  to reproduce the $^{3}$H beta decay and the binding energy of $^{4}$He.  With this choice, the binding energy of $^{3}$H is -8.32~MeV to be compared to the experimental value of -8.48~MeV.
When we perform the SRG transformation of the \emph{sole} NN-N3LO interaction we already obtain evolved NN+3N interactions. We will refer to this as the ``induced'' Hamiltonian. Conversely, the ``full'' Hamiltonian is the one obtained by also evolving the original 3NF-NNLO. Therefore, the effects of 3NFs of the Fujita-Miyazawa type are included in the full Hamiltonian only.

All calculations were performed in a model space of 12 harmonic oscillator (HO) shells \hbox{$[N_{max}\equiv\textrm{max}(2n+l)=11]$},  including all NN matrix elements and limiting 3NF ones to configurations with $N_{1}+N_{2}+N_{3}\leq N^{3NF}_{\textrm{max}}=14$.
 We checked that increasing $N^{3NF}_{\textrm{max}}$ from 14 to 16 changes Gorkov total binding energies by $\approx$500~keV. 
Changing the oscillator frequency between $\hbar\Omega_{HO}=$~20 and 24 MeV in Dyson-ADC(3) calculations, we found up to 450 KeV variations in the binding energy of $^{24}$O. Similarly, varying $\lambda_{\textrm{SRG}}$ in a limited range $1.88-2.0$ fm$^{-1}$ did not induce variations of more than $0.5\%$. From these and other tests we infer a conservative theoretical error of at most 5\%, for binding energies obtained with Dyson-ADC(3)~\cite{Cipollone2013prl}.
 Similar conclusions can be drawn about the prediction of dominant quasiparticle peaks in the single particle spectrum, $\e^+_k$ and~$\e^-_n$. Varying both $\hbar\Omega_{HO}=$~20-24~MeV and $\lambda_{\textrm{SRG}}=$~1.88-2.0~fm$^{-1}$,  we found a maximum variation of 310 KeV for the neutron 1/2$^-$ quasihole in $^{24}$O. This corresponds to  2\% of its value, $\e^-_{\nu1/2^-}=$~-14.22~MeV. The largest variation for proton quasiparticle energies was found to be of~550 KeV for a 5/2$^+$ quasiparticle, mostly due to variations in $\hbar\Omega_{HO}$. Therefore, we estimate theoretical errors of $\leq$1~MeV for the Dyson-ADC(3) gaps discussed below. For Gorkov calculations, we expect that errors on binding energies and quasiparticle peaks will be larger due to the simpler many-body truncation. However, we note that Ref.~\cite{Duguet2014ESPE} has reported a remarkable independence of dominant quasiparticle peaks on the $\lambda_{SRG}$ cutoff already at second order.


In the following sections, we will report the results obtained for $\hbar\Omega_{HO}$=~24~MeV and $\lambda_{SRG}$=~2~fm$^{-1}$.



\section{Results}
\label{results}

\begin{figure}[t]
\begin{center}
        \includegraphics[width=1.0\linewidth]{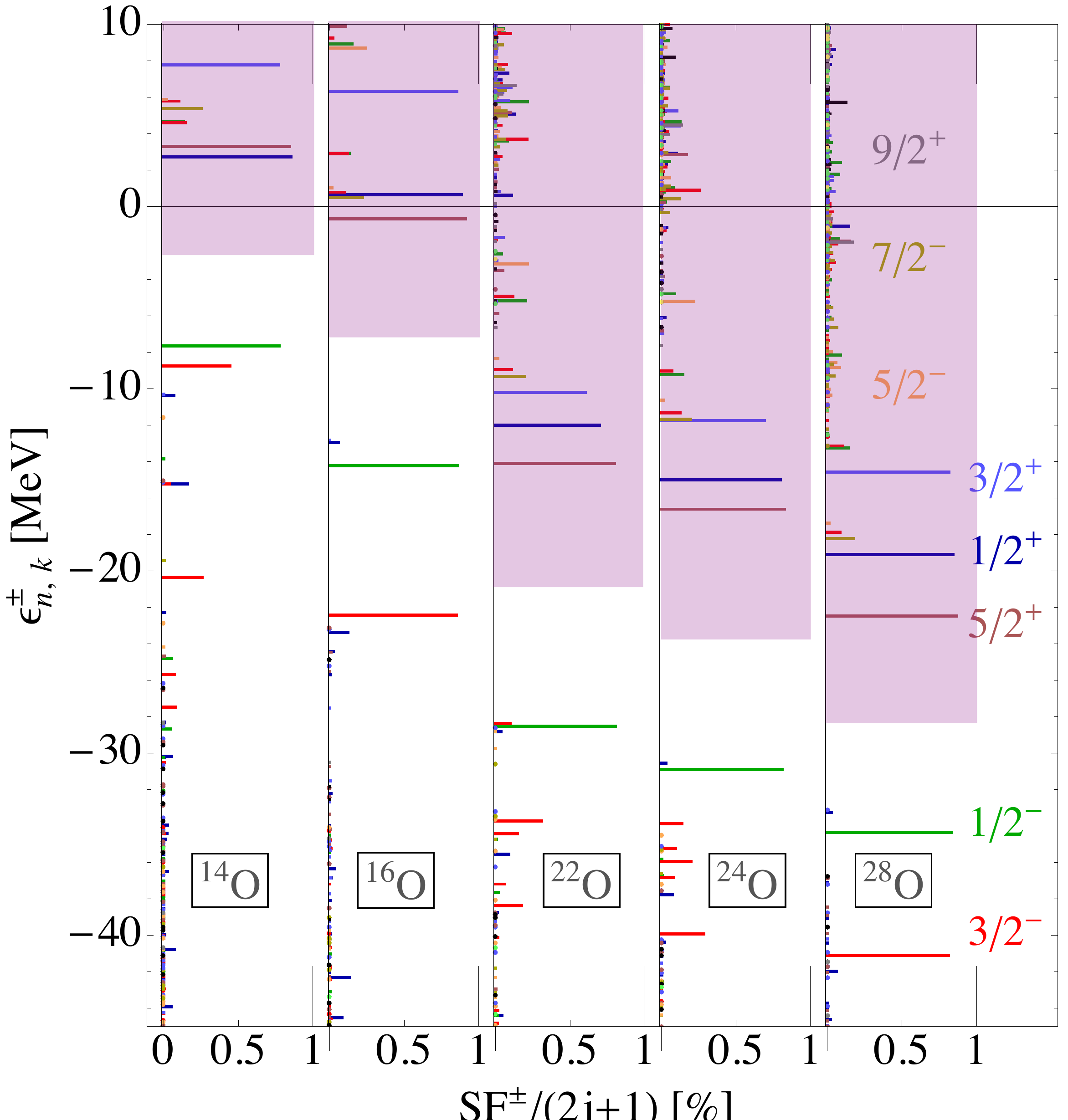}
	\caption{(Color online) Single particle spectral distributions for the addition and removal of a proton  to/from closed subshell oxygen isotopes. States above the Fermi Surface ($E_F$) are indicated by the shaded areas and yield the spectra of the resulting odd-even fluorine isotopes. The spectra below $E_F$ is for odd-even nitrogen isotopes in the final state (this appears inverted in the plot, with higher excitation energies pointing downward). Fragments with different angular momentum and parity are shown with different colors, as indicated, and the bar lengths provide the calculated spectroscopic factors. These results are obtained from ADC(3) and the full NN+3NF interaction with $\lambda_{SRG}=2.0$~fm$^{-1}$. } 
\label{fig:ADC3_SpectDist_prot}
\end{center}
\end{figure}

\begin{figure}[t]
\begin{center}
        \includegraphics[width=1.0\linewidth]{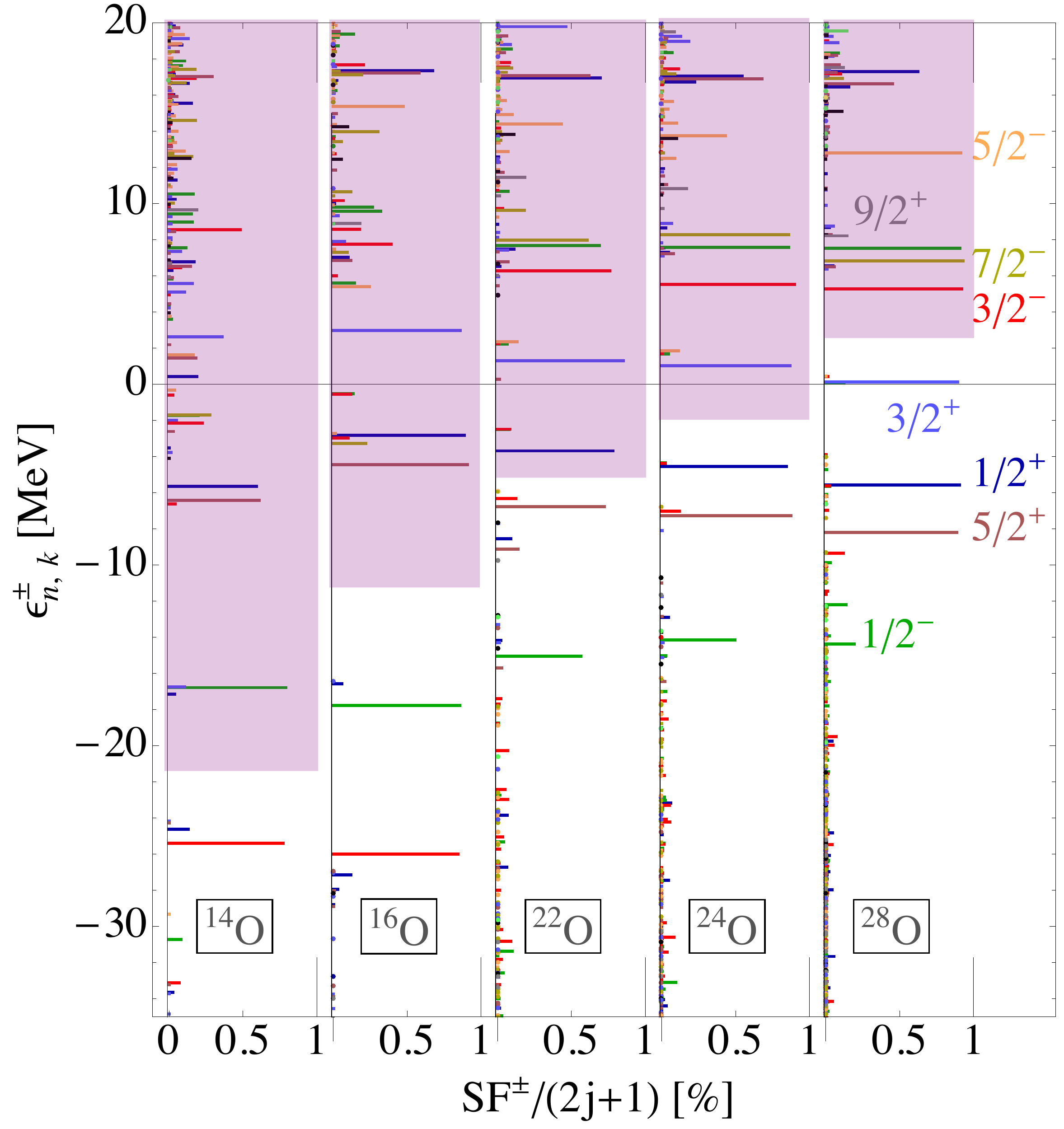}
	\caption{(Color online) Same as Fig.~\ref{fig:ADC3_SpectDist_prot} but for the addition and removal of a neutron. Both particle (shaded areas) and hole spectra are for odd-even oxygen isotopes.}
\label{fig:ADC3_SpectDist_neut}
\end{center}
\end{figure}

\subsection{Spectral functions and evolution of single particle spectra}

Three dimensional plots of the full spectral function, Eq.~\eqref{eq:SpSh}, are illustrated in Appendix~\ref{app_3D}. Here, we focus on the energy distribution of the spectral strength calculated by integrating its diagonal part over the single particle degrees of freedom,
\begin{eqnarray}
  S(\omega) &=& \sum_\alpha S^p_{\alpha \alpha}(\omega) + S^h_{\alpha \alpha}(\omega)  \nonumber \\
          &=& ~~ \sum_n  SF_n^+ \, \delta( \omega - E^{A+1}_n + E^A_0) \nonumber \\
          && +  \sum_n  SF_k^- \, \delta( \omega - E^A_0 + E^{A-1}_k ) \; ,
\label{eq:SFvsE}
\end{eqnarray}
which yields the energy distribution of spectroscopic factors. Each peak corresponds to eigenstate of a neighbouring odd-even isotope, whose energy is directly observed in nucleon addition and removal experiments.

The particle and hole contributions to Eq.~\eqref{eq:SFvsE}, calculated with Dyson ADC(3), are displayed in Fig.~\ref{fig:ADC3_SpectDist_prot} for protons and in Fig.~\ref{fig:ADC3_SpectDist_neut} for neutrons. 
The nucleon addition part of the spectra are highlighted by the shaded areas.
These figures show the general features of the correlated spectral distribution, which conserves strong quasiparticle fragments close the the Fermi surface but becomes heavily fragmented as one moves further away due to coupling to 2p1h and 2h1p (or more complex) excitations.  Quasiparticle states  with positive energies are above the one-nucleon continuum threshold ($E^{A+1}_n - E^A_0 $= 0 MeV) and therefore represent states for scattering of a nucleon off the $| \Psi^A_0 \rangle$ target. Since we assume a discrete model space in our calculations the associated particle continuum  is found discretised in several peaks that become more dense with increasing energy, reflecting the changes in the density of states for $| \Psi^{A+1}_n \rangle$.
Quasihole fragments at large negative energies correspond to highly excited $| \Psi^{A-1}_k \rangle$ states and also display a continuum portion of the spectrum. However, the spectral strength for nucleon removal is less pronounced. This due to the fewer degrees of freedom available to generate 2h1p configurations, and it can be equivalently explained in terms of the small overlap between the (A-1)-nucleon wave functions in the continuum and the bound $| \Psi^A_0 \rangle$ ground state.

\begin{figure}[t]
\begin{center}
        \includegraphics[width=1.0\linewidth]{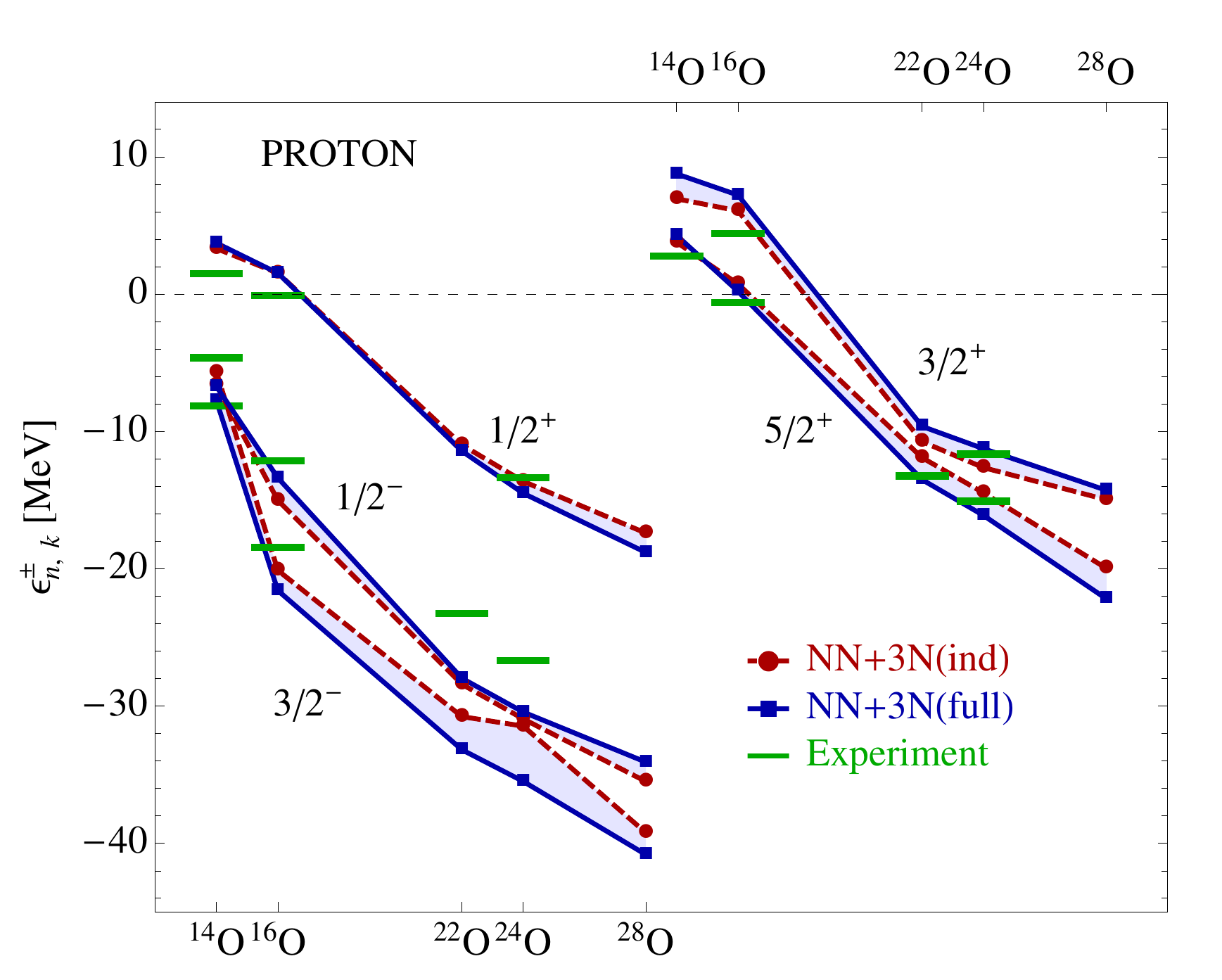}
	\caption{(Color online) Energy evolution of dominant proton quasiparticle fragments around oxygen isotopes of increasing neutron numbers.
	These level are for protons in the $p$ and $sd$ shells and refer to the ground and excited states of odd-even nitrogen (1/2$^-$, 3/2$^-$) and fluorine (5/2$^+$, 1/2$^+$, 3/2$^+$).
	Dots (joined by dashed lines) shows ADC(3) results obtained with the induced NN+3NF  interaction.
	 Squares (with full lines) refer to the full Hamiltonian, with the leading NNLO-3NF included. 
	 The latter are the dominant peaks also displayed in Fig.~\ref{fig:ADC3_SpectDist_prot}. 
	 In all cases $\lambda_{SRG}=2.0$~fm$^{-1}$.
	 Experimental values are from Refs.~\cite{AME2003,Jurado2007,Gaudefroy2012,Vajta2014F25ex}. }
\label{fig:ADC3_pks_prot}
\end{center}
\end{figure}

\begin{figure}[h!]
\begin{center}
        \includegraphics[width=1.0\linewidth]{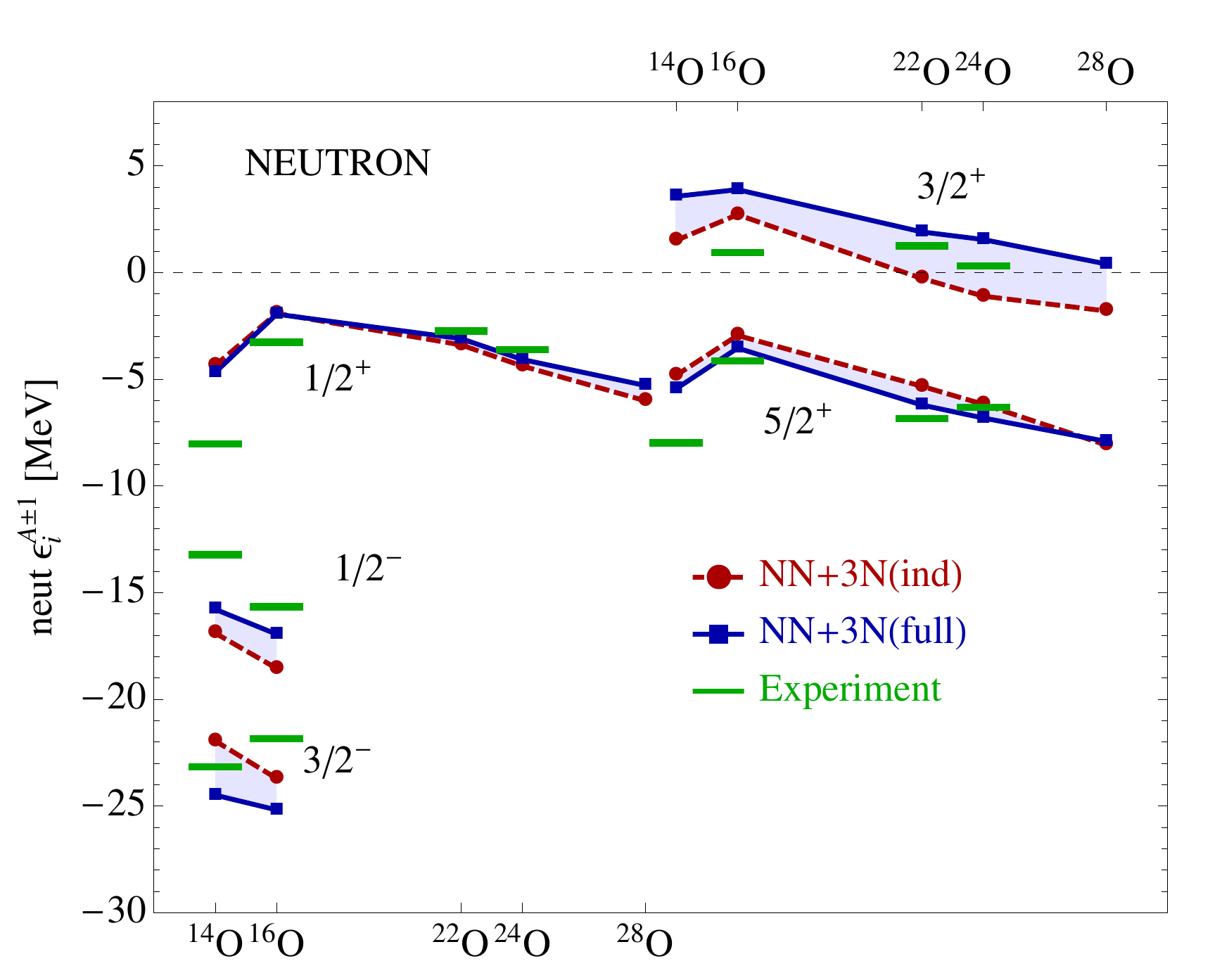}
	\caption{(Color online) Same as Fig.~\ref{fig:ADC3_pks_prot} but for dominant neutron fragments. Dots (joined by dashed lines) shows ADC(3) results obtained with the induced NN+3NF interaction. Squares (with full lines) refer to the full Hamiltonian and the same shown in Fig.~\ref{fig:ADC3_SpectDist_neut}.
	Experimental values are from Refs.~\cite{AME2003,Jurado2007,Hoffman2008prlO25,Kanungo2009prl,Elekes2007prlO23,Caesar2013R3BO26}.
	}
\label{fig:ADC3_pks_neut}
\end{center}
\end{figure}

\begin{table*}[t]
\centering
\begin{tabular}{lcccccccccccccccccccc}
\hline
\hline
 $\varepsilon_{dLS} \equiv$        &~&  \multicolumn{9}{c}{protons}  &~\quad ~&   \multicolumn{9}{c}{neutrons} \\
 \; $\varepsilon_{3/2^+} - \varepsilon_{5/2^+}$
        &~&  $^{15}$F    &~&  $^{17}$F    &~&  $^{23}$F    &~&  $^{25}$F    &~&  $^{29}$F 
        &~&  $^{15}$O   &~&  $^{17}$O   &~&  $^{23,21}$O   &~&  $^{25,23}$O    &~&  $^{27}$O \\
\\ \hline \\
$\Delta\varepsilon_{dLS}$: 
                                                &~& 1.32   &~&   1.67   &~&   2.72   &~&  3.02   &~&  2.92   &~&   2.70   &~&   1.77  &~&  3.06   &~&  3.30  &~&  2.03 \\
$\varepsilon_{dLS}^{IND}$:      &  &  3.17  &  &   5.33   &  &   1.17   &  &  1.84   &  &  4.98  &~&   6.33    &~&   5.65  &~&  5.05   &~&  5.06  &~& 6.28 \\
$\varepsilon_{dLS}^{FULL}$:    &  & 4.48   &  &   7.00   &  &  3.88   &  &  4.86    &  &  7.90   &  &  9.02    &  &   7.42  &  &  8.12   &  &  8.36  &  &  8.32 \\
$\varepsilon_{dLS}^{exp.}$:      &  &           &  &   5.00   &  &            &  &  3.83/3.44   &  &        & &           &  &  5.09   &  &  8.10   &  &    6.64   \\     
\hline
\end{tabular}
\caption{Spin-orbit splittings between 3/2$^+$ and 5/2$^+$ quasiparticle fragments. 
   For the case of protons these are eigenstates of the odd-even $^{A+1}$F isotopes indicated in the second row.
   In the case of neutrons, they are states of odd-even $^{A\pm1}$O. Note that for neutrons and A=22,24, these two levels are found
   across the Fermi surface and correspond to eigenstates of different isotopes.
   Results are reported for both the NN+3N-induced  and full Hamiltonians and
   $\Delta\varepsilon_{dLS} = \varepsilon_{dLS}^{FULL}$-$\varepsilon_{dLS}^{IND}$ are the 
   changes due to adding the original 3NFs at NNLO. Experimental values are from Refs.~\cite{AME2003,Jurado2007,Hoffman2008prlO25,Kanungo2009prl,Elekes2007prlO23,Caesar2013R3BO26,Vajta2014F25ex}.
}
\label{tab:dLS}
\end{table*} 

\begin{table}[t]
\centering
\begin{tabular}{lcccccccccccccc}
\hline
\hline
$\varepsilon_{pLS} \equiv $       &  \multicolumn{9}{c}{protons}  &\quad~&   \multicolumn{3}{c}{neutrons} \\
 \; $\varepsilon_{1/2^-} - \varepsilon_{3/2^-}$
        &  $^{13}$N    &~&  $^{15}$N    &~&  $^{21}$N    &~&  $^{23}$N    &~&  $^{27}$N 
        &~&  $^{15,13}$O   &~&  $^{15}$O \\
\\ \hline \\
$\Delta\varepsilon_{pLS}$: 
                                               &    2.00   &~&  3.10  &~&  2.84  &~&  4.52  &~&  3.01  &  &   3.66    &  &   3.01 \\
$\varepsilon_{pLS}^{IND}$:     &  -0.92  &~&  5.10  &~&  2.36  &~&  0.53  &~&  3.73  &  &   5.05    &  &   5.16  \\
$\varepsilon_{pLS}^{FULL}$:  &   1.07   &  &  8.21  &  &  5.20  &  &  5.05  &  &  6.74  &  &   8.71    &  &   8.24  \\
$\varepsilon_{pLS}^{exp.}$:    &   3.50   &  &  6.32  &  &           &  &           &  &           &  &   9.95    &  &   6.18  \\     
\\
\hline
\end{tabular}
\caption{Spin-orbit splitting between  1/2$^-$ and 3/2$^-$ quasiparticle fragments. 
   For the case of protons these are eigenstates of the odd-even $^{A-1}$N  isotopes indicated in the second row.
   For neutrons, they are states of odd-even $^{A-1}$O, except for $^{14}$O where the two 
   quasiparticles correspond to different isotopes.
   Results are reported for both the NN+3N-induced  and full Hamiltonians and
   $\Delta\varepsilon_{pLS} = \varepsilon_{pLS}^{FULL}$-$\varepsilon_{pLS}^{IND}$ are the
   contributions due to original 3NFs at NNLO.
   Experimental values are from Refs.~\cite{AME2003,Gaudefroy2012}.
}
\label{tab:pLS}
\end{table} 

\begin{table}[t]
\centering
\begin{tabular}{lcccccccccccccc}
\hline
\hline
$E_{gap} \equiv $       &  \multicolumn{9}{c}{protons}  &\quad~&   \multicolumn{3}{c}{neutrons} \\
 ~ $\varepsilon^{A+1}_{5/2^+}$
        &  $^{14}$O    &~&  $^{16}$O    &~&  $^{22}$O    &~&  $^{24}$O    &~&  $^{28}$O 
        &~&  $^{14}$O   &~&  $^{16}$O \\
  $ - \varepsilon^{A-1}_{1/2^-}$ & \\
   \hline \\
$\Delta E_{gap}$: 
                               &  0.61    &~&   -2.16   &~&   -2.03  &~&  -2.14   &~&  -3.64  &~&  -1.74   &~&   -2.20 \\
$E_{gap}^{IND}$:     & 10.38   &~&   15.76  &~&  16.50  &~&  16.46  &~&  15.54  &~& 12.07   &~&  15.60 \\
$E_{gap}^{FULL}$:  &  10.99   &  &   13.60  &  &  14.47  &  &  14.32  &  &  11.90 &  &  10.32   &  &  13.40 \\
$E_{gap}^{exp.}$:    &    7.41   &  &   11.53  &  &  10.02  &  &  13.33  &   &           &  &   5.24   &  &   11.52   \\     
  \\     
\hline
\end{tabular}
\caption{Energy gaps between the dominant 5/2$^+$ and 1/2$^-$ quasiparticles. These give
   a measure of the gaps between the $sd$ and $p$ shells for the $^A$O isotopes indicated in
   the second row.
   For the case of protons, these are particle-hole gaps and coincide with the ground states of the
   corresponding odd-even $^{A+1}$F and  $^{A-1}$N isotopes. 
   For neutrons, these are eigenstates of odd-even $^{A\pm1}$O but are not necessarily situated
   across the Fermi surface.
   Results are reported for both the NN+3N-induced  and full Hamiltonians and
   $\Delta E_{gap} = E_{gap}^{FULL}-E_{gap}^{IND}$ are the
   effects of the original 3NFs at NNLO. Experimental data are from Refs.~\cite{AME2003,Jurado2007,Hoffman2008prlO25,Kanungo2009prl,Gaudefroy2012,Caesar2013R3BO26,Vajta2014F25ex}.
}
\label{tab:Egap}
\end{table}

The fragments of the spectral distribution provide the excitation spectrum for the neighbouring odd-even isotopes. For example, the two dominant quasihole peaks in $^{24}$O, in Fig.~\ref{fig:ADC3_SpectDist_neut}, correspond to the 1/2$^+$ ground state and the 5/2$^+$ excitation of $^{23}$O. Our calculated 
excitation energy for the 5/2$^+$ state is 2.74~MeV, close to the experimental value of 2.79(13)~MeV~\cite{Schiller2007prlO23}. The 3/2$^+$ state of $^{23}$O can be calculated from the quasiparticle spectra of $^{22}$O. For this we obtain 5.0~MeV excitation energy, which is larger than the experimental value of 4.0~MeV~\cite{Elekes2007prlO23}.
In both cases, the theoretical result agrees with the {\em ab initio} configuration interaction (CI) calculations of Refs.~\cite{Bogner2014prlOxSM,Jansen2014prlOxSM}, which use the same NN+3NF full Hamiltonian. 
As already mentioned above, satellite peaks (that is, non dominant ones) are not necessarily well described in nucleon-attached and nucleon-removal methods at the ADC(3) level. This because they require leading order configurations of 2p1h/2h1p type or higher. The first 1/2$^+$ exited state of $^{21}$O, seen as a hole on $^{22}$O, is of this type and has a spectroscopic factor $\approx$9\% of the independent particle model. In spite of this, the ADC(3) excitation energy is 1.78~MeV which is again in great agreement with CI calculations based on the same Hamiltoninan (and slightly off the experimental value of 1.22~MeV~\cite{Stanoiu2004Oex}). Instead, the calculated spectroscopic factor the the 3/2$^+$ excited state is only $<$1\% and this is unlikely to be converged with respect to the many-body truncation in the ADC(3). For this state, we obtain an excitation energy of 940~keV that disagrees with both the experiment and the {\em ab initio} CI results, as expected. These results give a further confirmation of the performance of the present chiral Hamiltonian with the single $sd$ shell. Furthermore we note that the comparison with Refs.~\cite{Bogner2014prlOxSM,Jansen2014prlOxSM}  provides a successful benchmark of the accuracy of ADC(3) for calculating dominant quasiparticle states. We then use the latter to discuss the single particle structure across both $p$ and $sd$ shells.

Figure~\ref{fig:ADC3_pks_prot} shows the details of the evolution of the dominant proton quasiparticle and quasihole peaks, in the $sd$ and $p$ shells, for increasing neutron number. These are corrected for the effects of the COM motion according to Eqs.~\eqref{eq:com}. The dashed lines are obtained from the NN+3N-induced  interaction and represent the spectrum predicted by the initial N3LO two-nucleon force.  In general, the addition of original 3NFs (full lines) has the effect of consistently increasing the spin-orbit splittings between the 1/2$^-$--3/2$^-$ and the 3/2$^+$--5/2$^+$ dominant peaks. The $s_{1/2}$ orbit remain largely unaffected. The overall changes introduced by leading order 3NFs are reported in Tabs.~\ref{tab:dLS} and~\ref{tab:pLS} for both protons and neutrons.
The evolution of quasiparticle energies for the addition and the removal of a neutron is displayed in Fig.~\ref{fig:ADC3_pks_neut}. In this case, the $1/2^-$ and $3/2^-$ strength (in the $p$ shell) is strongly fragmented for masses above A=20 and no clear dominant peak is predicted. The original 3NFs still have the effect of increasing the splitting between spin-orbit partner states. However, this is in addition to the stronger repulsion on the $d_{3/2}$ orbit that is at the origin of the anomalous dripline at $^{24}$O~\cite{Otsuka2010prl}.

Worth of mention are the splittings between the 1/2$^-$ and the 3/2$^-$ quasiholes in $^{16}$O. For protons, this is predicted to be 5.1~MeV by the NN+3N-induced interaction, which is close to the empirical value of 6.32~MeV. However, the full Hamiltonian increases it to 8.2~MeV, overestimating the experiment. Exactly the same situation is found for the splitting between corresponding neutron holes, which is also increased by 3.1~MeV due to the original 3NF at NNLO. 
 For comparison, the Argonne v18 interaction that has a strongly repulsive core predicts a separation of $\approx$3.1~MeV~\cite{Barbieri2006plbO16} for these two states, at the NN interaction level. The corresponding Urbana~IX 3NF increases this by another 2.7~MeV predicting a splitting that is much closer to the experiment~\cite{Pieper1993prlN15}. Both three-nucleon Hamiltonians include two-pion terms of the Fujita-Miyazawa type and it is therefore reasonable that they generate similar corrections.

 From Tabs.~\ref{tab:dLS} and~\ref{tab:pLS} it is clear that the present NN+3N chiral Hamiltonians have a slight tendency to stretch the single particle spectrum, as compared to the experimentally observed dominant peaks. Corrections to this flaws may come at the price of introducing extra short-range repulsion in the NN interactions, for example through higher chiral cutoffs. At lower resolution scales, this implies the possible presence of relevant many-body forces at least at the 4NF level.
Ref.~\cite{Soma2014s2n} pointed out that the experimental gaps between the $sd$ and $pf$ shells are over estimated for the Ca and neighbouring isotopic chains. To investigate the behaviour in the present case, we consider the separations between the dominant $5/2^+$ and $1/2^-$ fragments that is representative of the gap between the $p$ and $sd$ shells.
These are reported in Tab.~\ref{tab:Egap}. With the only exception of the proton gap in $^{14}$O, we find that pre-existing 3NFs have the effect of reducing the distance between the two shells, by about 2~MeV, and to bring it closer to the experiment. In spite of this, the gaps remain consistently predicted too large by just a few MeV even when the full Hamiltonian is used.

\begin{figure}[t]
\begin{center}
   \includegraphics[width=1.0\linewidth]{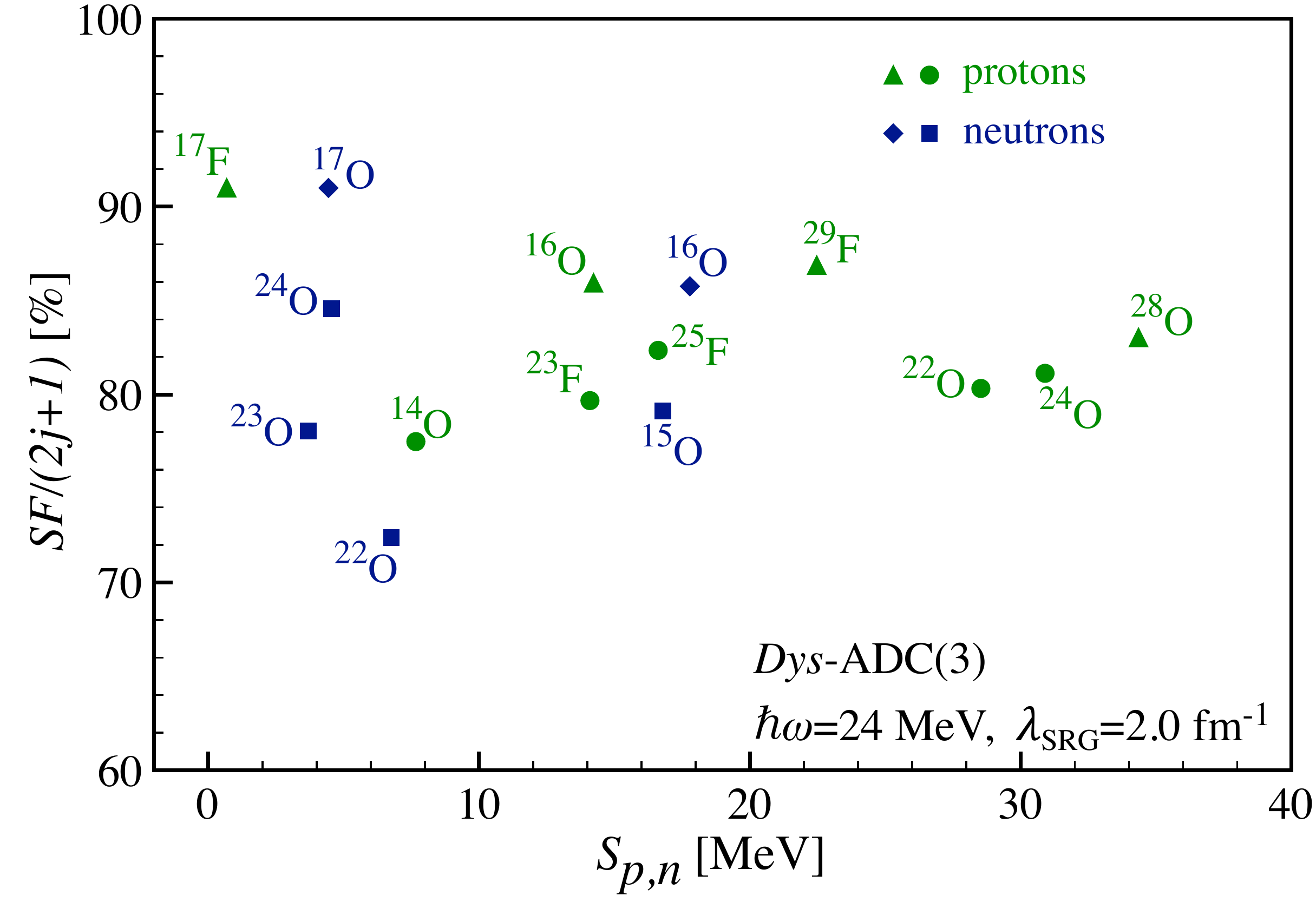}
   \caption{(Color online) 
      Calculated spectroscopic factors plotted as a function of their separation energies. 
      All values shown are for transitions between two ground states and refer to the removal
      of a proton (blue squares and diamonds) or a neutron (green circles and triangles) from
      the isotope indicated nearby. Triangles and diamonds highlight transitions that involve
      the $^{16}$O and $^{28}$O isotopes, at major shell closures.
      These results are for ADC(3) and the full NN+3NF interaction. 
      }
\label{fig:SFs}
\end{center}
\end{figure}

\subsection{Spectroscopic factors}

  The quenching of spectroscopic factors (SFs) for the dominant quasiparticle peaks can
provide useful insights on the strength of the correlations generated by 
the Hamiltonian.
 The principal mechanisms that are responsible for these, are identified with the coupling
of nucleons to high-momentum states due to short-range physics and with long-range effects
that include collective resonance modes and configuration mixing at small excitation
energies~\cite{Barbieri2009prl}.
For short-range correlations, we refer to the effects of the repulsive part of the 
central and  tensor NN interactions, typically at distances $<$1~fm, traditionally
used to reproduce nuclear phase shifts at very high energies.
Even for Hamiltonians that present strongly repulsive NN cores, the effects of 
short-range physics is usually found to be at most a 10\%~reduction
with respect to the independent particle model prediction~\cite{Dickhoff2004ppnp}. 
Thus, the quenching of SFs is mostly a consequence of low-energy physics. 
For the present Hamiltonian, the SRG evolution completely removes any quenching
due to short-range correlations. 

Since spectroscopic factors are directly linked to the cross sections
probed by particle addition and removal processes, it has long been debated
whether their evolution 
with proton-neutron asymmetry can explain the observed variations 
in the strength of direct nucleon knockout cross sections~\cite{Gade2008sf,Lee2010prlAr}.
The difference between the proton and neutron separation energies is
normally taken as a measure of such asymmetry.
A case of particular interest is $^{14}$O because of the very large value of this quantity.
The present ADC(3) calculations yield substantially the same spectroscopic factors equal
to 77.4\%~(77.2\%) for the removal of a proton (neutron) from this isotope to the ground
state of $^{13}$N~($^{13}$O).
Recent measurements of the (d,$^3$He) and (d,$^3$H) reactions are found
to be consistent with our calculations and therefore support a near independence
of correlations effects from proton-neutron asymmetry~\cite{Flavigny2013}.

In order to extend the analysis to cases with larger differences between proton
and neutron numbers, we plot in Fig.~\eqref{fig:SFs} the SFs for ground state
to ground state transitions along the whole chain. In general, we find values
evenly spread between 70\% and 90\% of the independent particle model.
 The smaller values of SFs are obtained at low separation energies
and involve transitions to/from $^{14,22,24}$O. These isotopes present reduced particle-hole
neutron gaps and therefore allow for stronger correlations at the Fermi surface.
 This consideration is also consistent with previous
works that clearly showed a close correlations between the particle-hole gap at the Fermi
surface and the predicted values of SFs~\cite{Barbieri2009ni}. From this, one may
infer that the over streched spectra reported in Tabs.~\ref{tab:dLS}, \ref{tab:pLS}
and~\ref{tab:Egap} result in more modest quenchings of SFs than otherwise expected.

 By looking only at transitions that involve the doubly-closed major shells $^{16}$O
and $^{28}$O, one can still identify a correlation between SFs and nucleon separation
energies. In particular, proton orbits tend to be more deeply bound as the number of neutrons
increases. This is due to the strong components of the proton-neutron forces, which also
enhances their correlations.
 However, the overall dependence on proton-neutron asymmetry is rather mild. We note
that the vicinity to the neutron dripline would require to  explicitly account for the continuum.
Ref.~\cite{Jensen2011prlOxSF} found that this effect is sizeable for $^{24,28}$O and leads
to further quenching of the proton SFs. 
Again, this could be interpreted as a reduced
gap between the highest neutron quasihole state and the nearby particle continuum. In this
sense, the reduction of spectroscopic factors is an indirect consequence of the change in
proton-neutron asymmetry, which first affects energy gaps.

For the case of the NN+3N-induced Hamiltonian we find a completely similar picture, with SFs
of dominant peaks being on average slightly larger than those obtained with the 
full interaction. Also in this case, stronger quenchings are associated with increased 
fragmentation of nearby strength and the narrowing of \hbox{(sub-)shell} gaps.
Thus, we conclude that the general effects of the original 3NFs on the quenching
of absolute SFs  mainly results from the rearrangement of shell orbits and 
excitation gaps.

\subsection{Results for open shells}

\begin{figure}[t]
\begin{center}
   \includegraphics[width=1.0\linewidth]{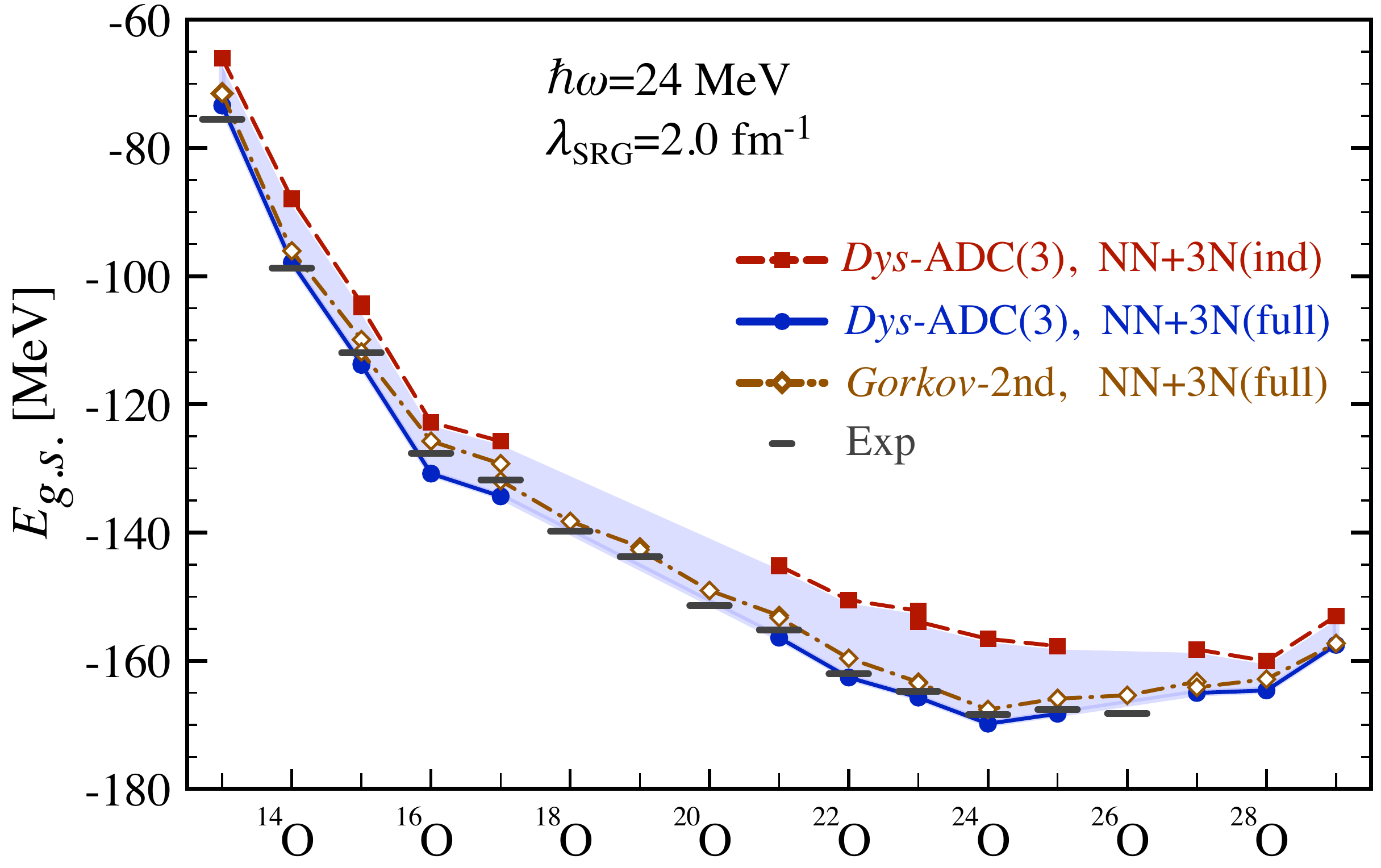}
   \caption{(Color online) Binding energies of oxygen isotopes. Dashed and full lines join the results
   from Dyson-ADC(3) calculations with the  NN+3N-induced (squares) and full (circles) Hamiltonians.
   The shaded area highlights the changes due to the original 3NF at NNLO.
   The open diamonds, joined by dot-dashed lines, are from Gorkov calculations at second order and include
   open shell isotopes.  Odd-even isotopes are obtained by summing total binging energies of the even-even
   systems, Eq.~\eqref{eq:Koltun_hW}, and the energies for addition or removal of a neutron, Eq.~\eqref{eq:com}. 
   Experiment are from Refs.~\cite{AME2003,Lunderberg2012prlO26,Caesar2013R3BO26,Jurado2007,Hoffman2008prlO25}.
 }
\label{fig:Gkv_BE_Ox}
\end{center}
\end{figure}


The present implementation of the Gorkov-GF approach allows calculations up
to the second order in the self-energy [i.e. at the ADC(2) level].  
Although this does not guarantee the best precision for quasiparticle
energies~\cite{vonNiessen1984CPR}, it still yields proper
predictions for the trend of binding energies~\cite{Soma2014s2n}. 
 
We plot the Gorkov predicted binding energies for all even-even isotopes in
Fig.~\ref{fig:Gkv_BE_Ox} and compare them to the Dyson-ADC(3) results where
available.
 For the Dyson case, the  NN+3N-induced Hamiltonian systematically under binds the full 
isotopic chain and predicts $^{28}$O to be bound with respect to $^{24}$O.  This is fully
corrected by including the original 3NF at leading order, which brings all results
to about~3\% form the experiment  or closer. This is well within the estimated
theoretical errors discussed above~\cite{Cipollone2013prl}.
 The dot-dashed line shows the trend of ground state energies for the full Hamiltonian 
obtained form Gorkov, which include the $^{18,20,26}$O isotopes. This demonstrates
that the fraction of binding  missed by the second order truncation is
rather constant across the whole isotopic chain and, in the present case, of
about 2-4~MeV. The result is a constant shift with respect to the complete ADC(3) 
prediction and the overall trend of binding energy is reproduced very close
to the experiment.
Note that binding energies for odd-even oxygens can be calculated either as
 neutron addition or  neutron removal from two different nearby isotopes.
Fig.~\ref{fig:Gkv_BE_Ox} shows that this procedure can lead to somewhat different
results, which should be taken as an indication of the errors due to the second
order many-bod truncation. For the more complete Dyson-ADC(3) method and
the full Hamiltonian, this differences are never larger than 200~keV and are not
visible in the plot.

\begin{figure}[t!]
\begin{center}
   \includegraphics[width=1.0\linewidth]{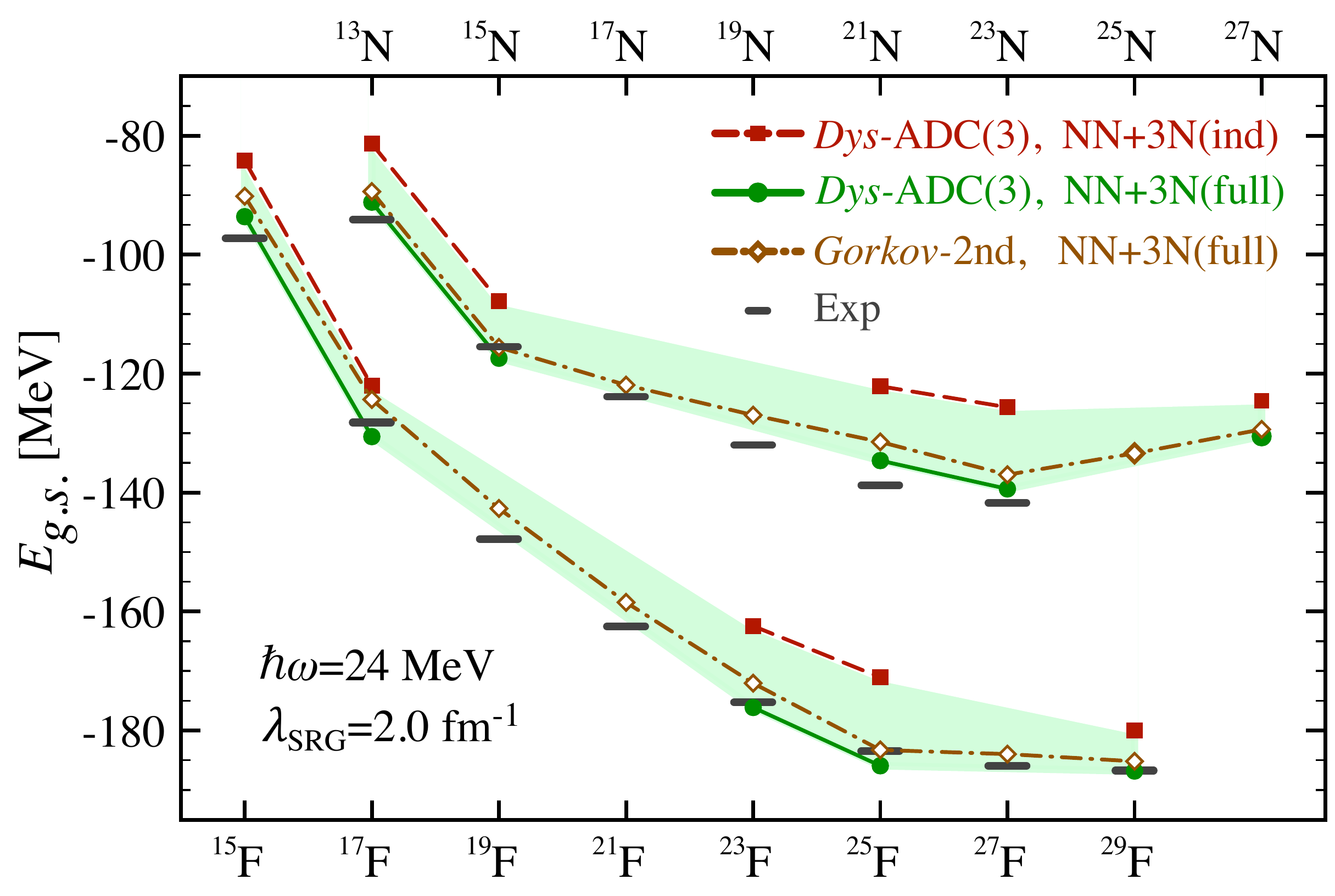}
   \caption{(Color online) Same as Fig.~\ref{fig:Gkv_BE_Ox} but for the binding energies of nitrogen
   and fluorine isotopes. These are calculated as addition or removal of a proton to and from
   even-even oxygen isotopes.
   Experiment are from Refs.~\cite{AME2003,Lunderberg2012prlO26,Caesar2013R3BO26,Jurado2007,Gaudefroy2012}.
 }
\label{fig:Gkv_BE_FN}
\end{center}
\end{figure}

Figure~\ref{fig:Gkv_BE_FN} shows the analogous information for the binding energies
of the nitrogen and fluorine isotopic chains, obtained through removal and addition
of one proton. This confirms that all considerations made regarding the effects
of leading order 3NFs on the oxygens also apply to their neighbouring chains. In particular,
the repulsive effect on the $d_{3/2}$ neutron orbit is key in determining the neutron
driplines at $^{23}$N and $^{24}$O. Fluorine isotopes have been observed experimentally
up to $^{31}$F but with a $^{29}$F that is very weakly bound.  Fig.~\ref{fig:Gkv_BE_FN} 
clearly demonstrates that this is due to an very subtle cancellation between the repulsion
form 3NFs and the attraction generated by one extra proton~\cite{Cipollone2013prl}.
Our calculations with the more accurate Dyson-ADC(3) scheme predict $^{28}$O
to be unbound with respect to $^{24}$O by 5.2~MeV. However, this value should
be slightly affected by the vicinity to the continuum~\cite{Hagen2012prlOx}, which
was neglected in the present work.

The general qualitative features of the spectral functions discussed
in the previous sections are also found in our Gorkov propagators but 
with an even more spread single particle spectra. 
For example, the splitting among the 1/2$^-$ and 3/2$^-$ quasihole
states of $^{15}$N is found to be 10.2~MeV, compared to the 8.2~MeV
calculated in the Dyson-ADC(3) scheme [cfr. Tab.~\ref{tab:pLS}].
This larger value is a consequence of neglecting the interactions
between 2p1h and 2h1p configurations by the second order truncation.
Interestingly, this splitting is sensibly reduced in the neighbouring
semi-magic isotopes and it is calculated to be 4.9~MeV
for  $^{17}$N and 5.6~MeV for $^{19}$N. These  
values refer to the separation from the first $3/2^-$ state close to the
Fermi surface (rather than a centroid of the first few fragments). They are
sensibly smaller because the calculations yield a fragmented $p_{3/2}$ hole orbit
for these nuclei.
  In the Gorkov calculations, the changes in these splitting due 
to aiding the original NNLO-3NF ($\Delta\varepsilon_{pLS}$) are  4.2, 0.8 and 2.2~MeV,
respectively for  $^{15}$N, $^{17}$N and~$^{19}$N.


\begin{figure}[t]
\begin{center}
        \includegraphics[width=1.0\linewidth]{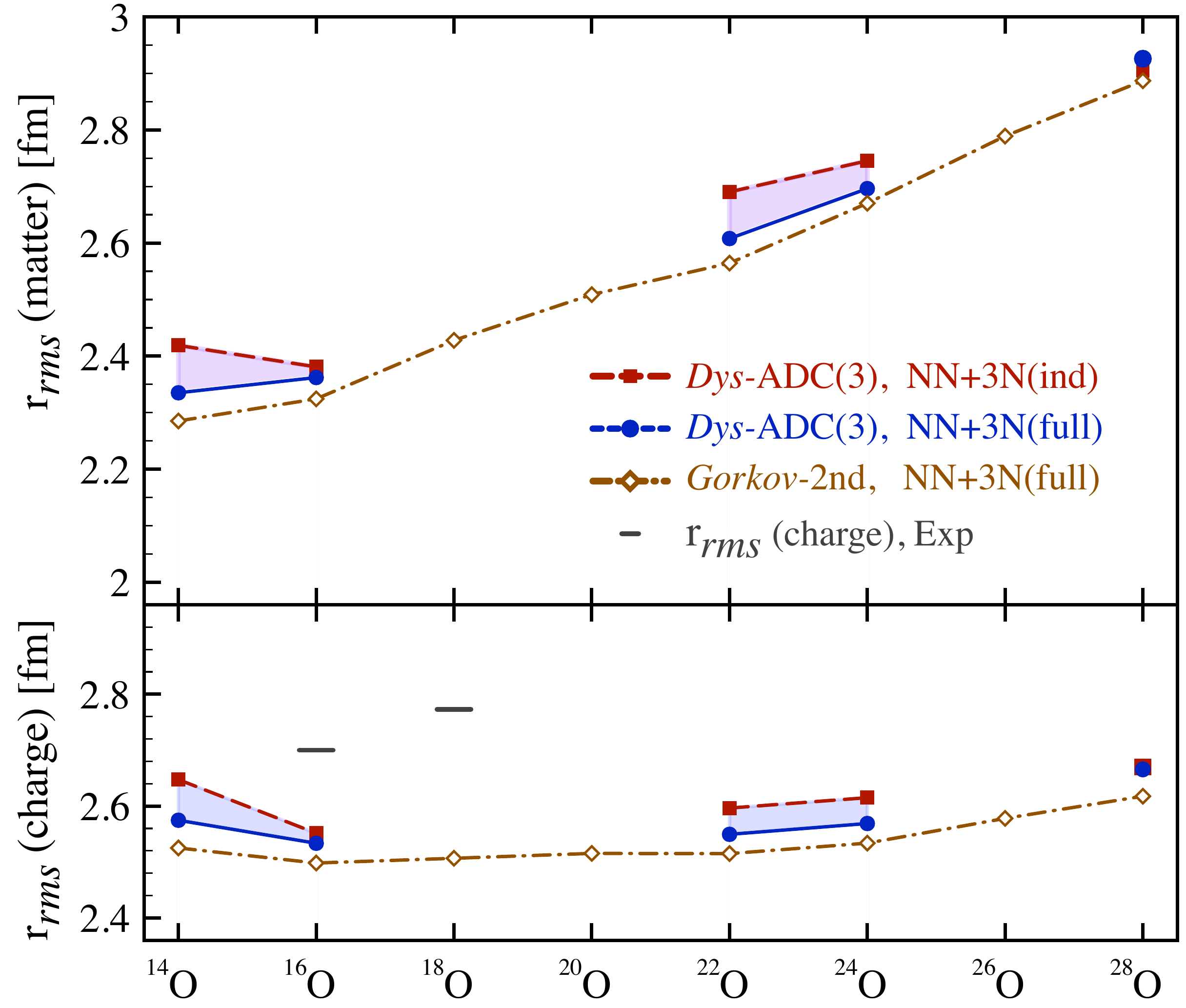}
	\caption{(Color online) Matter radii (top panel) and charge radii (bottom) obtained from Dyson-ADC(3) and second order Gorkov.
	 Curves are the same as in Fig.~\ref{fig:Gkv_BE_Ox}. 
	 The bars are experimental charge radii from Ref.~\cite{Angeli2013ADNDT}.
	 }
\label{fig:radii}
\end{center}
\end{figure}

Figure~\ref{fig:radii} demonstrates the trend obtained for point-matter and charge radii along
the whole chain and compares them to the observed charge radii for $^{16}$O and~$^{18}$O.
 The induced 3NFs give sizeable contributions to the calculated radii, which 
would be sensibly smaller if computed from the evolved NN-only interaction~\cite{Barbieri:2012rd}.
Conversely, the original 3NFs cause only a small reduction. Eventually,
the radii predicted by the complete Dyson-ADC(3) calculations and the full Hamiltonian
are smaller than the experimental charge radii by 0.2-0.3~fm.
 Note that the present calculations do not account for the evolution of operators 
through the SRG. On the other hand, the small radii are consistent with the over stretched 
spectra discussed above.
 The radii calculated with the second order Gorkov approach give somewhat
smaller results due to the many-body truncation.  Nevertheless, they describe
the overall trend of increasing matter radii along the whole chain, while charge
radii remain largely independent of neutron number.
Note that the available experimental data shows almost equal charge radii for $^{16}$O
and $^{17}$O and a slight larger value for $^{18}$O~\cite{Angeli2013ADNDT}.
This is reminiscent of the behaviour of isotope shifts in calcium isotopes that first
increase  and then return small moving when going from $^{40}$Ca and  $^{48}$Ca, although 
more data up to $^{22}$O would be required to confirm this. The bell shape
observed in calcium isotopes is explained by extensive shell model calculations
covering both $sd$ and $pf$ shells~\cite{Caurier20012CaIsoShf,Avgoulea2011ScRch}.
Such type of correlations are not included in the present Gorkov formalism
at second order and therefore the flat behaviour of charge radii of Fig.~\ref{fig:radii}
is consistent with the many-body truncations adopted here.

\section{Conclusions}
\label{conlusions}

We have presented a comprehensive study of the single particle spectral functions and
ground state properties of oxygen isotopes based on chiral NN+3N interactions. 
To this goal, we performed {\em ab initio} calculations within self-consistent Green's function~(SCGF)
theory. The theoretical framework of this approach has been reviewed highlighting
the physics information contained in the nuclear spectral function.
Calculations were performed for the closed subshell isotopes using the Dyson-ADC(3)
many-body truncation scheme which is presently the state of the art technique. For
the  open shell isotopes $^{18}$O, $^{20}$O and $^{26}$O we performed calculations 
using the recently introduced Grokov formulation of SCGF, which can be currently
applied at second order.

The general features of the nuclear spectral functions have been discussed, with
particular emphasis on the quasiparticle energies (also referred to as ``separation'' energies)
for the dominant peaks observed in the removal and addition of a nucleon. The
3N interactions at NNLO have the effect of increasing the spin orbit splittings
of the $p$ and $d$ orbits and  lead to over estimating the experiment.
At the same time, the 3NFs reduce the gaps between the $p$ and $sd$ major shells, improving
the agreement with data but not enough to reproduce the empirical values. 
We observe that all these deficiencies might be corrected by having extra short-range
repulsion in the NN section of the Hamiltonian.
Other approaches, such as global fittings of chiral NN+3N forces to include  
medium mass isotopes, also hold the promise to reach proper saturation~\cite{Ekstrom2015NNLOsat}.
In general, it is found that the current NN-N3LO interaction with cutoff at 500~MeV,
augmented  by the local  3NF-NNLO with a 400~MeV local cutoff, tends to stretch the single
particle spectrum too much compared to data. The corresponding predictions
for matter radii under estimate the experiment. 

The conclusion that the present chiral forces overestimate the gaps between
major shells was already pointed out in Ref.~\cite{Soma2014s2n} for isotopic
chains around Ca, and suggest that these saturate nuclear matter at slightly
higher densities than the empirical point~\cite{Carbone2013snm,Carbone2014corr3nf}.
  Here, we find that hints of the same pathologies are seen also for the oxygen
isotopes, in spite of the fact that binding energies are nicely predicted at smaller
masses.

 The calculated absolute quenchings of spectroscopic factors
change only mildly with proton-neutron asymmetry. This is valid as long as
the occupied states in the single particle spectrum are not near to the continuum.
Stronger correlations would instead be generated by smaller particle-hole gaps
in the oxygen isotopes with closed subshells.

\acknowledgments

The authors would like to thank T. Duguet and V. Som\`a for their collaboration on developing Gorkov-SCGF codes. 
This work was supported by the United Kingdom Science and Technology Facilities Council (STFC) under Grants Nos. ST/J000051/1, ST/L005743/1 and ST/L005816/1 and by the Natural Sciences and Engineering Research Council of Canada (NSERC), Grant No. 401945-2011.  Calculations were performed using HPC resources from the DiRAC Data Analytic system at the University of Cambridge (BIS National E-infrastructure capital grant No. ST/K001590/1 and STFC grants No. ST/H008861/1, ST/H00887X/1, and ST/K00333X/1).
TRIUMF receives funding via a contribution through the Canadian National Research Council.\\

\appendix

\begin{figure}[h!]
\begin{center}
        \includegraphics[width=0.97\linewidth]{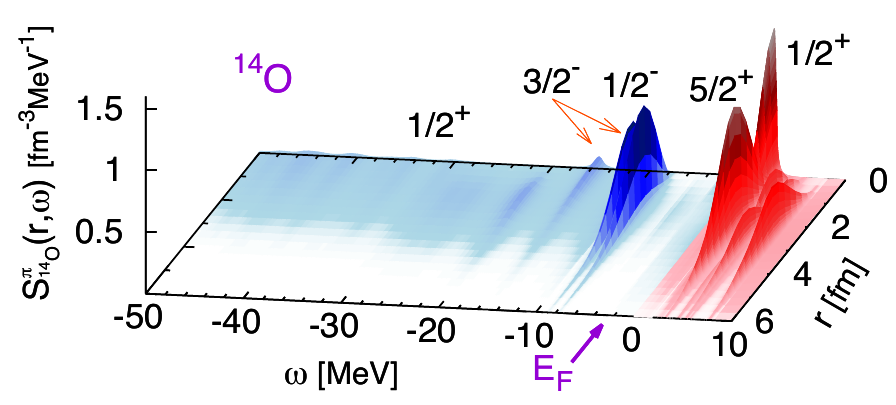}
	\vskip .1cm
        \includegraphics[width=0.97\linewidth]{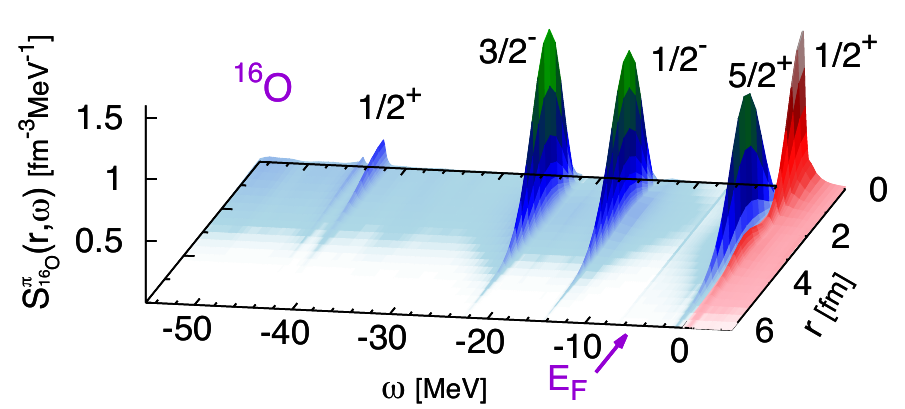}
	\vskip .1cm
        \includegraphics[width=0.97\linewidth]{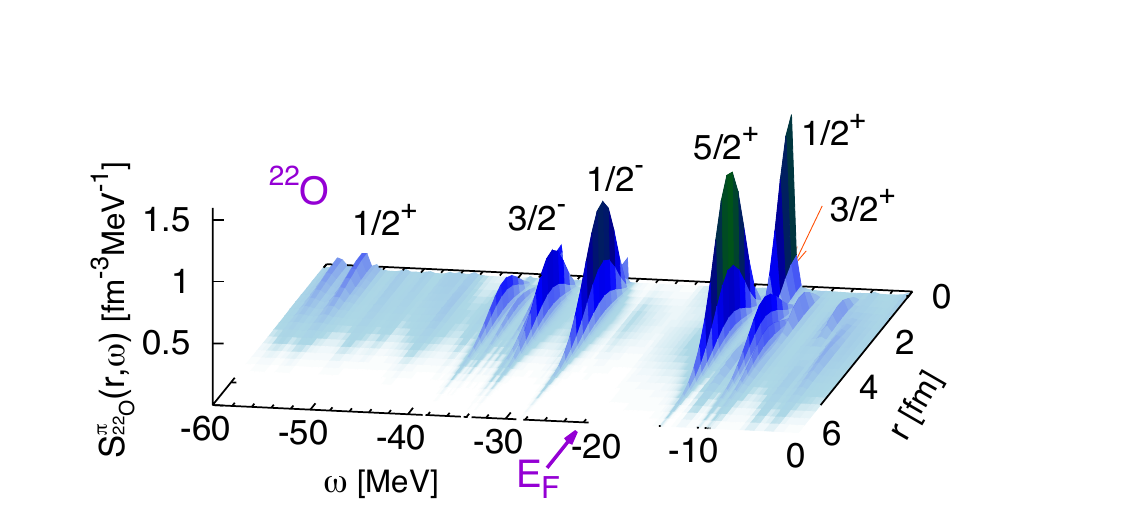}
	\vskip .1cm
        \includegraphics[width=0.97\linewidth]{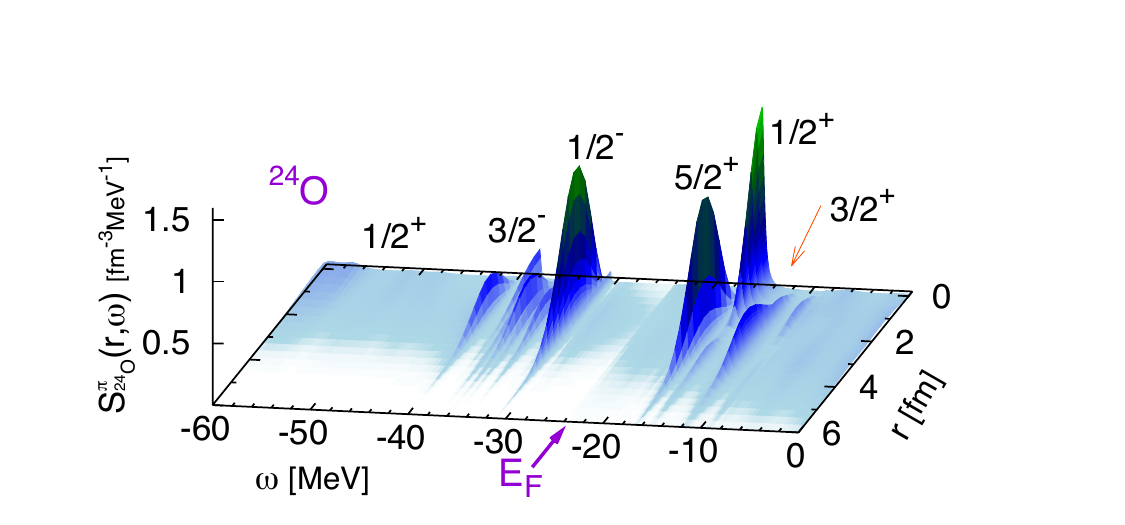}
	\vskip .1cm
        \includegraphics[width=0.97\linewidth]{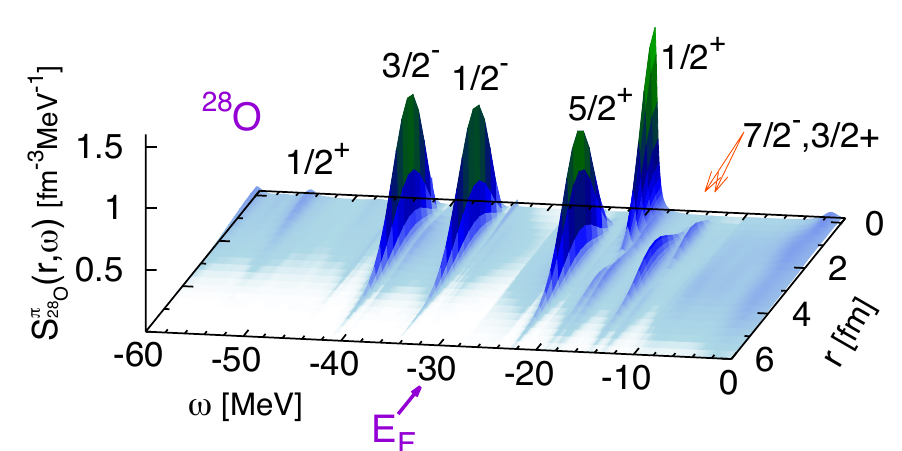}
	\caption{(Color online) Diagonal part of the complete proton spectral function, Eq.~\eqref{eq:S_of_r_w},
	for closed subshell isotopes $^{14,16,22,24,28}$O. The discretised energy peaks that appear as energy delta
	functions in Eq.~\eqref{eq:SpSh} have been smeared with Lorentzians of suitable with.
	Energies below the Fermi surface, $E_F$, correspond to the hole part of the spectral distribution while those above are for particle addition. 
	Energies $\omega>$~0~MeV (plotted in red) correspond to proton-nucleus scattering states.
	}
	\vskip -1.4cm
\label{fig:S_3D_prot}
\end{center}
\end{figure}

\begin{figure}[h]
\begin{center}
        \includegraphics[width=0.97\linewidth]{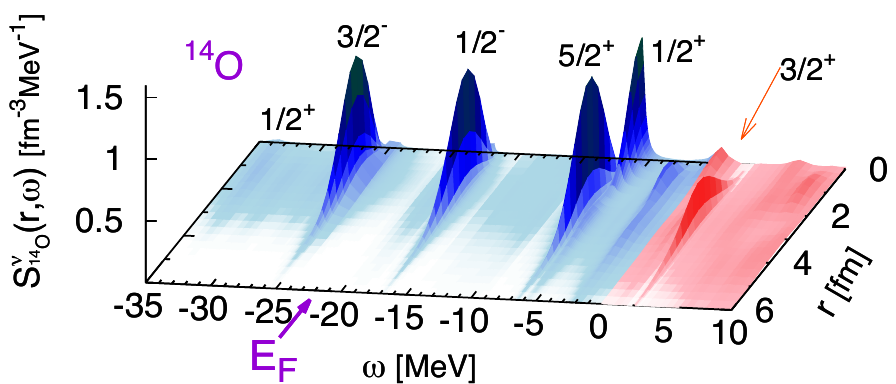}
	\vskip .1cm
        \includegraphics[width=0.97\linewidth]{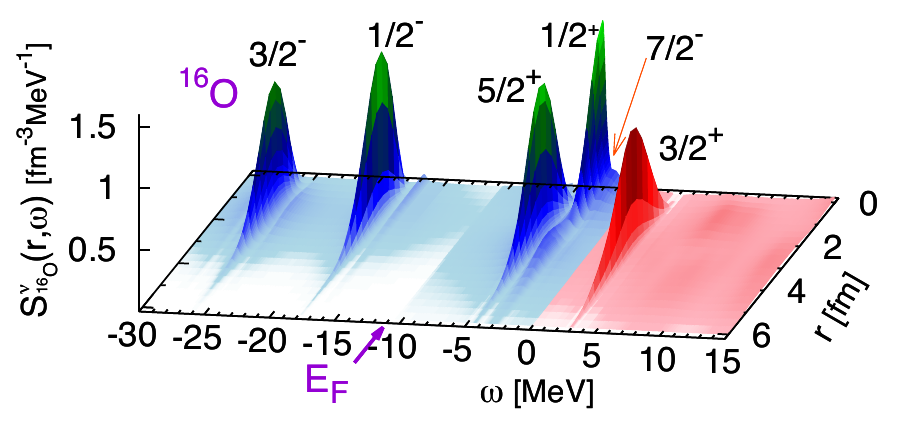}
	\vskip .2cm
        \includegraphics[width=0.97\linewidth]{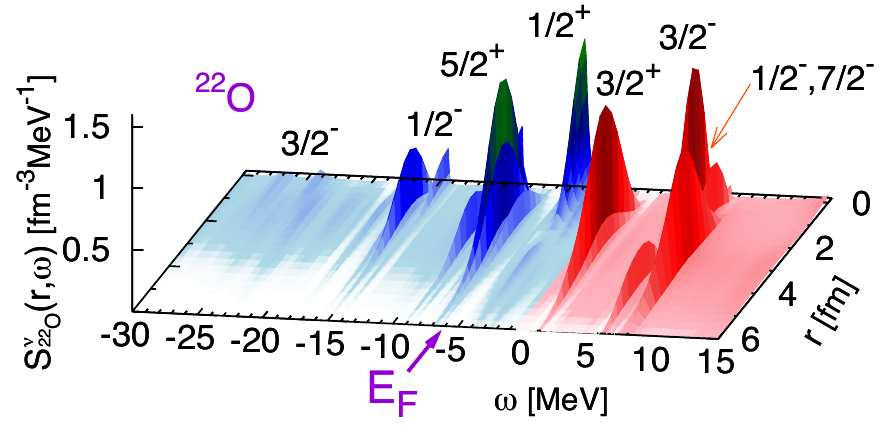}
	\vskip .1cm
        \includegraphics[width=0.97\linewidth]{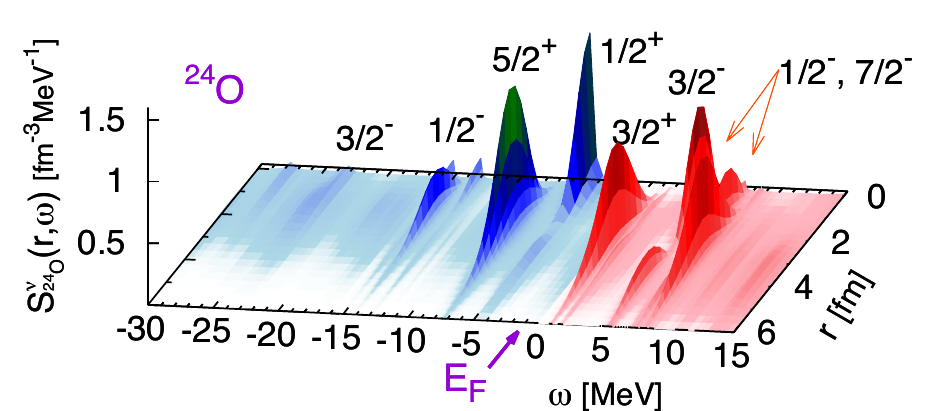}
	\vskip .2cm
        \includegraphics[width=0.97\linewidth]{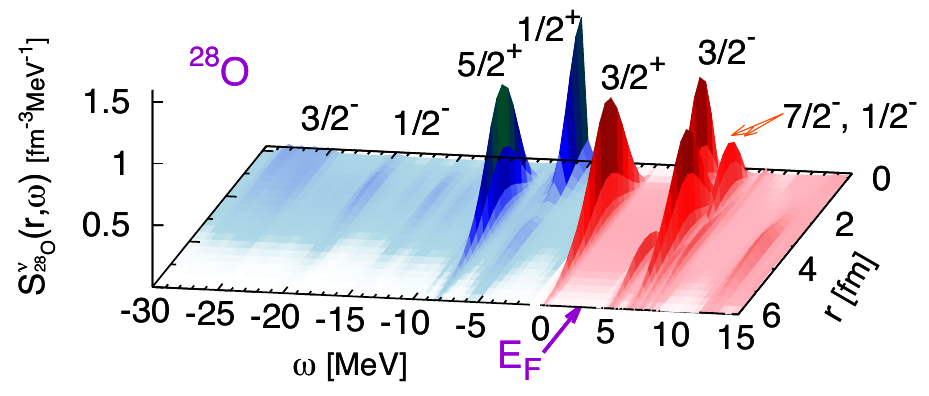}
	\caption{(Color online) 
	 Same as Fig.~\ref{fig:S_3D_prot} but for neutrons.}
	\vskip -1cm
\label{fig:S_3D_neut}
\end{center}
\end{figure}

\section{plots of spectral functions}
\label{app_3D}

The diagonal part of the one-nucleon spectral function, Eq.~\eqref{eq:SpSh}, has a straightforward physical interpretation~\cite{Fett1971book,Dickhoff:book}. 
Its particle part, $S^p_{\alpha \alpha}(\omega)$, is the joint probability of adding a nucleon with quantum numbers $\alpha$ to the A-body ground state, $|\Psi^A_0\rangle$, and then to find the system in a final state with energy $E^{A+1}=E^A_0 + \omega$. Likewise,  $S^h_{\alpha \alpha}(\omega)$ gives the probability of removing a particle from state $\alpha$ and later finding the nucleus in an eigenfunction of energy $E^{A-1}=E^A_0 - \omega$.
Once transformed to coordinate or momentum representations, these distributions give a rather intuitive picture of the single particle structure of a nucleus. 
We demonstrate this by calculating the spectral function in coordinate basis as
\begin{equation}
S(r, \omega) = \sum_{\alpha \; \beta} \phi_\alpha({\bf r}) \left[ S^p_{\alpha \beta}(\omega) + S^h_{\alpha \beta}(\omega) \right] \phi^*_\beta({\bf r}) \; ,
\label{eq:S_of_r_w}
\end{equation}
where $ \phi_\alpha({\bf r})$ are spin-coupled harmonic oscillator functions. The angular and spin dependences in Eq.~\eqref{eq:S_of_r_w} are removed by  summing over all oscillator sates. This appens for all nuclei under consideration because they have $J^\pi=$0$^+$ quantum numbers in their ground sates.

 The  spectral functions obtained from Dyson-ADC(3) calculations are displayed in Figs.~\ref{fig:S_3D_prot} and~\ref{fig:S_3D_neut} for protons and neutrons, respectively.  These shows the radial distribution of the squared one-nucleon overlap wave functions at different quasiparticle energies. The Fermi energy, $E_F\equiv(\varepsilon^+_0 + \varepsilon^-_0)/2$, marks the separation between the hole and particle parts of the spectral distribution. Hence, integration over all energies in the range $\omega\in[-\infty, E_F]$ yields the nucleon density $\rho(r)$ [see Eq.~\eqref{eq:rho}] and further integrations over coordinate space yields the particle number. 
  Note that quasiparticle states for $\omega>0$ correspond to the continuum spectrum of the corresponding (A+1)-nucleon system. These are unbound states for the  scattering of a nucleon off the $|\Psi^A_0\rangle$ ground state~\cite{Capuzzi96,Escher2002opt}. Thus, they extend to infinity in the limit of a complete single particle model space. In the present work, we only calculate their projection on a truncated harmonic oscillator space, which cannot be normalised to the usual asymptotic boundary conditions. Nevertheless, the plots put in evidence the predicted location for neutron resonances in the $sd$ and $pf$ shells. 
 It must be kept in mind that these resonances will be further corrected in extended calculations that properly account for the continuum. In general these effects will be more important the broader is the resonance
 and for the present case one may expect corrections as large as a few~MeV.
  Importantly, the self-energy, Eq.~\eqref{eq:ADCn}, is a bound function which can be correctly expanded even in an harmonic oscillator basis. Thus, by first transforming this to coordinate or momentum space, it is possible to obtain a complete optical potential and to compute scattering waves with proper boundary conditions. This is normally done in applications of SCGF theory to scattering~\cite{Waldecker2011frpadom,Barbieri:2005NAscatt}.


\bibliography{./SCGF_biblio}

\end{document}